\documentclass[preprint,12pt, showkeys]{revtex4}

\usepackage[brazil, english]{babel}
\usepackage[utf8]{inputenc}
\usepackage{graphicx}
\usepackage{epsfig}
\usepackage{amssymb}
\usepackage{amsmath} 
\usepackage{slashed}
\usepackage{xcolor}
\usepackage[unicode=true,bookmarks=false,breaklinks=false,pdfborder={0 0 1},colorlinks=true]
 {hyperref}
\hypersetup{
 citecolor=blue,linkcolor=blue,urlcolor=blue}

\graphicspath{%
    {converted_graphics/}
    {/}
}
\begin{document}

\title{Pion form factor from an AdS deformed background} 
\author{Miguel Angel Martín Contreras$^1$}
\email{miguelangel.martin@uv.cl}
\author{Eduardo Folco Capossoli$^{2}$}
\email{eduardo\_capossoli@cp2.g12.br} 
\author{Danning Li$^3$}
\email{lidanning@jnu.edu.cn}
\author{Alfredo Vega$^1$}
\email{alfredo.vega@uv.cl} 
\author{Henrique Boschi-Filho$^{4}$}
\email{boschi@if.ufrj.br}  
\affiliation{$^1$Instituto de Física y Astronomía, Universidad de Valpara\'iso, A. Gran Breta\~na 1111, Valpara\'iso, Chile\\ $^2$Departamento de F\'\i sica / Mestrado Profissional em Práticas de Educação Básica (MPPEB), Col\'egio Pedro II, 20.921-903 - Rio de Janeiro-RJ - Brazil\\ $^3$Department of Physics and Siyuan Laboratory, Jinan University, Guangzhou 510632, China\\ $^4$Instituto de F\'\i sica, Universidade Federal do Rio de Janeiro, 21.941-972 - Rio de Janeiro-RJ - Brazil}

\begin{abstract}
\textcolor{black}{We consider a bottom-up AdS/QCD model with a conformal exponential deformation  $e^{k_I\,z^2}$ on a Lorentz invariant AdS background, where $k_I$ stands for the scale $k_\pi$ that fixes confinement in the pion case and $k_\gamma$ for the kinematical energy scale associated with the virtual photon. In this model we assume the conformal dimension 
associated with the operator that creates pions at the boundary as $\Delta=3$, \textcolor{black}{as in the original bottom-up AdS/QCD proposals}. Regarding the \textcolor{black}{geometric slope} related to photon field $k_\gamma$, we analyze two cases: constant and depending on the transferred momentum $q$. In these two cases we computed the electromagnetic pion form factor as well as the pion radius. We compare our results with experimental data 
as well as other theoretical (holographic and non-holographic) models. In particular, for the \textcolor{black}{momentum dependent scale}, we find good agreement with the available experimental data as well as non-holographic models.} 
\end{abstract}


\maketitle


\section{Introduction}\label{intro}

\textcolor{black}{Since its proposal and its discovery, the pion still attracts a lot of attention from the high-energy community. The pion is the simplest hadron modeled by QCD and due to the two main features of QCD $-$ asymptotic freedom and confinement $-$ the pion itself plays a role of the most appropriate character to probe the interplay between soft and hard regimes. The soft regime is represented by low-energy scale (small $q^2$) and it can only be investigated by non-perturbative methods. On the other hand, hard regimes belong to the high-energy realm (large $q^2$) can be studied trough perturbative techniques.}
\textcolor{black}{Pions, along with nucleons, are non-perturbative particles responsible for nuclear stability since they are strong force mediators, and also they are the Goldstone bosons of QCD. In particular, the last property has a deeper connection with the mass generation in QCD. Thus, a proper understanding of the pion structure is needed. However, this issue is not accessible to address since its intrinsic non-perturbative nature.  Experimentally, the electromagnetic pion form factor was measured in the low $q^2$ region by scattering pions over electrons at CERN, as can be seen in Ref.} \cite{Amendolia:1986wj}.  \textcolor{black}{In the intermediate $q^2$ range (up to 5 MeV$^2$), the electromagnetic pion form factor is measured from the electroproduction of pions from nucleon scattering processes in facilities as  DESY, JLab, and Cornell, for instance in Refs.} \cite{Ackermann:1977rp, Tadevosyan:2007yd, Bebek:1977pe} \textcolor{black}{, respectively. Regarding theoretical aproaches this direction, one can find in literature, we many propositions to compute the pion form factor, such as,  QCD sum rules, dispersion relations with QCD constraint, Dyson-Schwinger equation, perturbative QCD (pQCD), light-front quark
model (LFQM), and extended vector meson dominance model (extVMD), and Lattice QCD at $N_f=2$. These approaches are presented in Refs.} \cite{Nesterenko:1982gc, Geshkenbein:1998gu, Maris:2000sk, Bakulev:2004cu, Choi:2006ha, Lomon:2016eyp, Frezzotti:2008dr} \textcolor{black}{, respectively.}

The end of the 1990s brought a major breakthrough in theoretical physics. The seminal papers written by Juan Maldacena \cite{Maldacena:1997re} and soon after by Witten \cite{Witten:1998qj, Witten:1998zw} and by Gubser, Klebanov, and Polyakov \cite{Gubser:1998bc} presented to us how to connect a $(d+1)$-dimensional gravitational theory to a $d$-dimensional conformal field theory (CFT). Such a relation is known as gauge/gravity correspondence or duality. On the gravity side the theory is established over a high-dimensional curved anti-de Sitter (AdS$_{d+1}$) spacetime (the bulk) meanwhile in the CFT side it is established over a $d$-dimensional flat Minkowski spacetime (the boundary). Although, conceptually speaking, this duality seems extremely abstract, it relationship with high-energy physics became quite straight forward since one can construct a dual gravitational theory in the bulk and related it to a super Yang-Mills theories on the boundary. In particular, and probably the main goal in its use, by breaking the conformal symmetry, one can study QCD-like theories or AdS/QCD models. 

The breaking of the conformal invariance is an important subject for the usage of AdS/CFT correspondence since real QCD is not a conformal theory. A very famous AdS/QCD model which breaks conformal symmetry and provides an excellent result for vector meson spectrocopy is the softwall model (SWM) as proposed in the very first time in Ref. \cite{Karch:2006pv} which includes a scalar (dilaton) field in the action for the fields.

However, it was argued in Refs. \cite{Li:2013oda, Capossoli:2015ywa} that the original formulation of the SWM, even breaking the conformal invariance, it seems to not work well for the glueball spectroscopy. Besides, in Refs. \cite{BallonBayona:2007qr, Braga:2011wa} it was also shown that the original formulation of SWM does not produce a mass gap for the fermionic sector. In order to overcome these mentioned questions one can see in Refs. \cite{Huang:2007fv, Branz:2010ub, Gutsche:2011vb, Afonin:2012jn, Fang:2016uer, Cortes:2017lgz, Afonin:2018era, Gutsche:2019blp, Contreras:2018hbi} an incomplete list where the authors proposed some modifications in the original SWM. Among many modifications, we rather follow the one proposed in Refs. \cite{Andreev:2006vy, Andreev:2006ct} where a scalar (dilaton) field is introduced in the AdS metric instead of in the action for the fields as done in the original formulation of SWM. 

By working within this approach the authors of Ref. \cite{FolcoCapossoli:2019imm} could, at same time, break the conformal invariance and compute compatible results for the masses of even and odd spin glueballs, scalar mesons, vector mesons and fermions with spins 1/2, 3/2 and 5/2. This approach was also used in the following incomplete list of Refs. \cite{Forkel:2007cm, White:2007tu, Wang:2009wx, Rinaldi:2017wdn, Bruni:2018dqm, Diles:2018wbe, Tahery:2020tub, Caldeira:2020sot, Chen:2020ath, Caldeira:2020rir, Rinaldi:2021dxh}, where one can check many works dealing with holographic high-energy physics and using some kind of deformation in the AdS space metric.

In this sense the AdS/QCD program is very suitable to study both soft and hard regimes at same time. Since the publication of the pioneer work in Ref. \cite{Grigoryan:2007wn} many authors have used some AdS/QCD model to study the pion \cite{Ferreira:2019inu, Lv:2018wfq, Bacchetta:2017vzh} or the pion form factor \cite{Grigoryan:2008up, Brodsky:2007hb, Kwee:2007dd, Vega:2008te, Bayona:2010bg, Zuo:2009hz, Stoffers:2011xe, Gutsche:2014zua}.

Here, in this work, by using an AdS/QCD model taking into account a deformed AdS$_5$ space, we will compute the pion electromagnetic form factor, which characterize the interaction of a pion with an external photon,  as a function $F_{\pi}(q^2)$ of the squared four-momentum transfer $q^2$. \textcolor{black}{Notice that  our calculation does not define the bulk field conformal dimension, $\Delta$, in terms of the meson constituent number with the twist. We keep the spectroscopic definition of  $\Delta$, i.e., defined in terms of the scaling dimension of the operators responsible for creating hadrons at the boundary. This choice implies that at large $q^2$ limit, pion form factors do not acquire the expected scaling behavior $1/q^2$. However, we propose a solution to this issue by promoting the virtual photon geometric slope to be depending on the transfer momentum. This solution fits well within these models, where each particle is defined with different backgrounds. The confinement scale, related to the pion, is different from the virtual photon scale associated with the kinematics of the scattering process.} \textcolor{violet}{Finally,} we compare our holographic results with the available experimental data as well as other theoretical holographic and non-holographic approaches.

This work is organized as follows: in Section \ref{defdis} we present our deformed AdS/QCD model as well as we describe the scalar and the gauge boson fields in the bulk which represent, respectively, the pion and the photon fields at the boundary. In Section \ref{pion}, after a brief review of the pion properties, we present the interaction action of the scalar and gauge fields and compute holographically the pion form factor. In Section \ref{numerical} we present our results in comparison with experimental data and other theoretical approaches. Finally, in Section \ref{conc} we present our conclusions.


\section{Deformed A\lowercase {d}S/QCD model}\label{defdis}

 In this section we present the deformed AdS/QCD model which will be used to calculate the pion form factor. Let us start writing its 5-dimensional action as:
\begin{equation}\label{acao_soft}
S = \int d^{5} x \sqrt{-g} \; {\cal L}\,, 
\end{equation}

\noindent where ${\cal L}$ is the Lagrangian density to be detailed below for the scalar and gauge fields, and $g$ is the determinant of the metric $g_{mn}$ of the deformed $AdS_5$ space:
\begin{equation}\label{metric} 
ds^2 = g^I_{mn} dx^m dx^n= \frac{e^{k_{I} z^2}}{z^2} \, (dz^2 + \eta_{\mu \nu}dy^\mu dy^\nu)\,, 
\end{equation}
\textcolor{black}{where the index $I=\pi, \gamma$ is associated with the pion and the photon, respectively. Note that the geometry $g^I_{mn}$ is different for each particle, since they interact differently with the static background. Accordingly, the parameters $k_\pi$ and $k_\gamma$ are related to the pion and the photon, respectively.}

To avoid possible misunderstandings one should note that we will use throughout this text the indices $m, n, \cdots$  to refer to the five-dimensional space, separating into $\mu, \nu, \cdots$ for the 4-dimensional Minkowski spacetime and the holographic $z$ coordinate. The Minkowski flat spacetime is endowed with metric $\eta_{\mu \nu}$ with signature $(-,+,+,+)$.

The introduction of the conformal exponential factor ${e^{k_I z^2}}$ in the AdS metric as above defines our deformed AdS space which is asymptotically AdS for $k_I \to 0$\textcolor{black}{. At the UV boundary $z\to 0$ the exponential deformation is negligible.}

\textcolor{black}{At this point it is very important to make a brief discussion in order to recall how confinement is achieved in our proposed model. When we are dealing with bottom-up models, we consider the duality between perturbative bulk fields living in an AdS space and non-perturbative operators at the boundary.  In the pure AdS, normalizable modes coming from the associated holographic potential form a continuum spectrum. Thus, confinement in these sorts of models is achieved by discretizing such a spectrum.  This procedure can be done by imposing a hard cutoff (as in the quantum square well) 
\cite{BoschiFilho:2002ta, BoschiFilho:2002vd} or modifying the holographic potential's large z behavior (Soft wall model and deformed background). The former defines the so-called hard-wall model. The latter opens the possibility to deform the geometry or use extra auxiliary bulk fields, as in the soft-wall model where a static dilaton field is used. In our particular case, we use a quadratic and static deformation function that induces linear Regge trajectories for baryons and mesons. It is in this frame where we set up our pion field, defined by the Regge slope $k_\pi$. 
In this case the deformation slope is  fixed to get the pion mass which defines the confining IR scale. 
This also implies that the geometry is not the same for the pion and for the virtual photon. The deformation slope in the case of the virtual photon, $k_\gamma$, is not associated with confinement. This parameter is related to the energy scale of the scattering process, as in the case of deep inelastic scattering in bottom-up models \cite{Polchinski:2002jw, Braga:2011wa, FolcoCapossoli:2020pks}.}

In the following subsections, we describe the free scalar and gauge fields actions, which represent, respectively, the pion and the photon fields. The interaction action of these fields which account for the electromagnetic pion form factor will be discussed in Section \ref{pion}.

\subsection{Scalar field in the deformed A\lowercase {d}S/QCD model}\label{secff}

A massive scalar field $X$ in the deformed AdS$_5$ space, Eq. \eqref{metric}, is described by the action:
\begin{equation}\label{esc_sw}
S = \int d^5 x \sqrt{-g_\pi}\; [ g^{mn}_\pi \partial_m X \partial^n X + M_5^2 X^2 ]\,,
\end{equation} 
\textcolor{black}{where $g^{\pi}_{mn}$ is the metric defined in Eq. \eqref{metric} related to the pion. }

It is worthwhile to mention that the scalar field in the bulk will represent the mesonic particles in 4 dimensions. In particular, we are interested in the pion particle.

From the action \eqref{esc_sw} one can find the following equation of motion:
\begin{equation} \label{eom_esc}
\partial_m[\sqrt{-g_{\pi}} g^{mn}_{\pi} \partial_n X] - \sqrt{-g_{\pi}}M_5^2 X = 0\,. 
\end{equation}
Writing $g^{mn}_\pi = e^{-2A_{\pi}(z)} \eta^{mn}$, with the warp factor $A_{\pi}(z)$ given by:
\begin{equation}
    A_{\pi}(z) = -\log z + \frac{k_\pi}{2}\, z^2\,,  \label{A}
\end{equation}
the equation of motion Eq.  \eqref{eom_esc} can then be rewritten as:
\begin{equation} \label{eom_esc_2}
\partial_m[e^{3A_{\pi}(z)} \eta^{mn} \partial_n X] - e^{5A_{\pi}(z)} M_5^2 X = 0, 
\end{equation}
or defining $B(z) = - 3A_{\pi}(z)$, one has:
\begin{equation}\label{eomsw2}
\partial_m[e^{-B(z)} ~ \eta^{mn}  \partial_n X] - e^{\frac{-5 B(z)}{3}} M^2_5 X = 0.
\end{equation}
Next, we use a plane wave ansatz with amplitude just depending on the $z$ coordinate and propagating in the transverse coordinates $x^{\mu}$ with momentum $q_{\mu}$, 
\begin{equation}\label{an}
X (z, x^{\mu}) = v(z)\, e^{-i q_{\mu} x^{\mu}}.
\end{equation}

\noindent After some algebraic manipulation and defining 
$v(z) = \psi (z) e^{\frac{B(z)}{2}}$ one has a ``Schr\"odinger-like'' equation:
\begin{equation}\label{eq_4}
- \psi''(z) + \left[ \frac{B'^2(z)}{4}  - \frac{B''(z)}{2} + e^{\frac{-2 B(z)}{3}} M^2_5 \right] \psi(z) = - q^2 \psi(z), 
\end{equation}
\noindent where $M_5$ is the scalar field (mesons) mass in five dimensions and $E = -q^2$ are the eigenenergies which represent the mesons masses in four dimensions.

By using the AdS/CFT prescription one can learn how  to  relate  the  bulk mass $M_5$ to the conformal dimension $\Delta$ of an operator in four dimensions, so that:
\begin{equation}\label{m5scalar}
   M^2_5 = (\Delta - p) (\Delta + p - 4)\,,
\end{equation}
\noindent where $p$ represents the index of the $p-$form which in this case is associated with the hadronic spin $S$. 


From the QCD description we know that the scalar mesons are composed by a bound state of quark-antiquark belonging to a spin singlet with total total angular momentum $J= L+S=0$ and for our purposes in this work we will disregard  all other meson quantum numbers. Besides scalar mesons are represented, in the boundary, by the operator:
\begin{equation}
{\cal O}_{SM}= \bar{q}\, D_{\lbrace J_1 \cdots} D_{J_m \rbrace}q \;\;\; {\rm with} \;\;\;\displaystyle\sum_{i=1} J_i = J
\end{equation}

The contribution coming from each quark is $3/2$, then the conformal dimension reads $\Delta = 3/2+ 3/2 =3$ and the bulk mass in Eq.\eqref{m5scalar} is $M_5^2=-3$. Replacing this result in Eq. \eqref{eq_4} one gets:
\begin{equation}\label{eq_sm}
- \psi''(z) + \left[ \frac{B'^2(z)}{4}  - \frac{B''(z)}{2} -3\; e^{\frac{-2 B(z)}{3}} \right] \psi(z) = - q^2 \psi(z)\,. 
\end{equation}
This equation does not have analytic solutions. Then,   solving it numerically with $k_{\pi}=-0.0425^2$ GeV$^2$ we get $m_{\pi}= 0.139$ GeV which is compatible with the meson $\pi$ mass \cite{pdg}. 

\begin{center}
\begin{figure}[h!]
  \begin{tabular}{c c}
    \includegraphics[width=3.2 in]{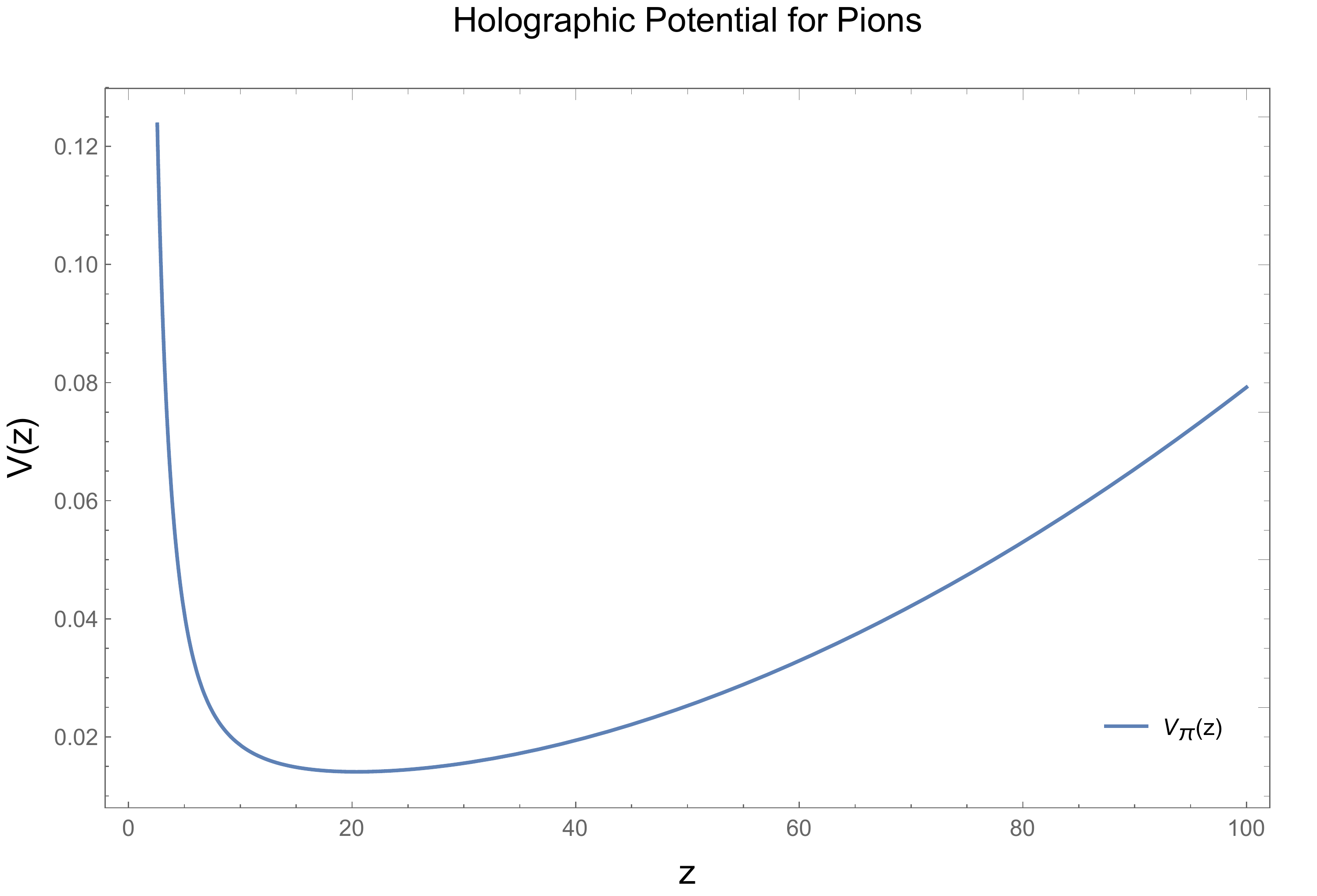}
    \includegraphics[width=3.2 in]{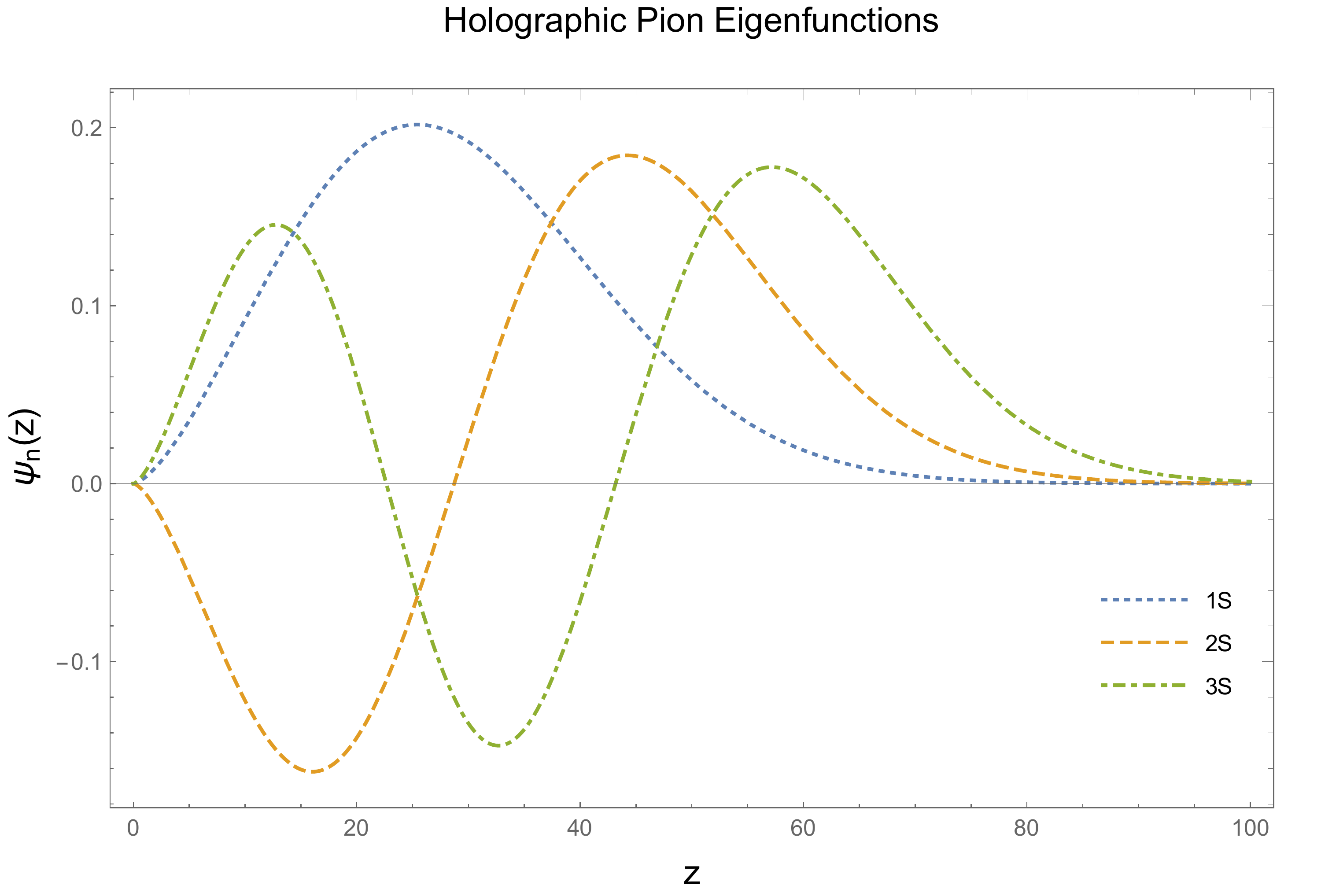}
  \end{tabular}

\caption{Left panel depicts the holographic potential for bulk eigenmodes dual to pions. Right panel shows the ground state and the first two excited bulk eigenmodes dual to pions.}
\label{fig:one}
\end{figure}
\end{center}

In Fig. \ref{fig:one} (left panel) we present the holographic potential for pions associated with the ``Schr\"odinger-like'' equation, Eq. \eqref{eq_sm}. In the right panel of Fig. \ref{fig:one} we present the holographic pion eigenfunctions for the states $1S$, $2S$ and $3S$.

\subsection{Gauge boson field in the deformed A\lowercase {d}S/QCD model}

Here in this section we  describe within our deformed AdS space the gauge boson field which represents the physical photon at UV. Such a photon will interact with the pion through an   electromagnetic current  contributing to the calculation of the pion form factor.

To do so we introduce the action for a five dimensional massless gauge boson field $\phi^n(x^{\mu}, z)$, so that:
\begin{equation}\label{f15}
S = - \frac{1}{c_\gamma^2} \int d^{5} x \sqrt{-g_{\gamma}} \; \frac{1}{4} F^{mn} F_{mn}\,,
\end{equation}
\noindent
\textcolor{black}{where $g^{\gamma}_{mn}$ is the metric presented in Eq. \eqref{metric} associated with the photon, } and the electromagnetic tensor $F^{mn}$ is written as usual as $F^{mn} = \partial^m \phi^n - \partial^n \phi^m$. 

From the above action one derives the equations of motion:
\begin{equation}\label{f16}
\partial_m [ \sqrt{-g_{\gamma}}\; F^{mn}] = 0\,. 
\end{equation}
Since the metric $g_{mn}^{\gamma}$, Eq. \eqref{metric}, is diagonal, for $n=z$ one gets
\begin{equation}\label{phiz}
 \Box \phi_z    - \partial_z [\partial_{\mu} \eta^{\mu \nu} \phi_{\nu}]= 0\,, 
\end{equation}
or simply
\begin{eqnarray}
\Box\,\phi_z-\partial_z\left(\partial_\mu\,\phi^\mu\right)=0\,, \label{solem2}
\end{eqnarray}
while for $n=\mu$ one has
\begin{eqnarray}\label{phinu}
e^{-A_\gamma(z)} \partial_{z}[e^{A_\gamma(z)} \partial_z \phi_{\nu}] +\Box \phi_{\nu}  -  \partial_{\nu} \left(e^{-A_\gamma(z)} \partial_{z}[e^{A_\gamma(z)}  \phi_{z}]+ \partial_{\alpha}  \phi^{\alpha}\right)=0\,,
\end{eqnarray}
\noindent where $\Box \equiv \eta^{\sigma \alpha} \partial_{\sigma} \partial_{\alpha}$.
The electromagnetic field profile comes from the solutions of Eqs. \eqref{solem2} and \eqref{phinu}. 
Choosing the gauge: 
\begin{equation}\label{gauge}
\partial_{\nu} \left(e^{-A_\gamma(z)} \partial_{z}[e^{A_\gamma(z)}  \phi_{z}]+ \partial_{\alpha}  \phi^{\alpha}\right)=0, 
\end{equation}
the Eq. \eqref{phinu}, written in Fourier space,  reduces to 
\begin{eqnarray} \label{solem}
-q^2\,\phi_\mu+A'_\gamma\,\partial_z\,\phi_\mu+\partial_z^2\,\phi_\mu=0\,, 
\end{eqnarray}

\noindent where $A_\gamma=A_\gamma(z)$ is analogous to Eq.\eqref{A} but now related to the photon, such as
\begin{equation}
    A_\gamma(z) = -\log z + \frac{k_{\gamma}}{2}\, z^2\,,  
\end{equation}

and prime denotes derivative with respect to $z$. 
This kind of gauge fixing was used in Refs. \cite{Polchinski:2002jw, Hatta:2007he, BallonBayona:2007qr, Braga:2011wa, Capossoli:2015sfa, FolcoCapossoli:2020pks}, when discussing deep inelastic scattering (DIS) holographically. Of course the gauge choice will not affect the physical results as, for instance, the pion form factor.

Further, we will consider a photon with a transversal  polarization $\eta_\mu$ such that $\eta_ \mu \, q^ \mu =0$. In this sense Eq. \eqref{solem2} will not contribute for our calculations, and only  the electromagnetic field component $\phi^\mu$ will be relevant to the pion form factor. Such a consideration was also done in Refs. \cite{Polchinski:2002jw, Hatta:2007he, BallonBayona:2007qr, Braga:2011wa, Capossoli:2015sfa, FolcoCapossoli:2020pks}, in holographic DIS studies to calculate structure functions.

The general solution to equation \eqref{solem} has the following form: 
\begin{eqnarray}
\phi_\mu(z,q)=C^1_{\mu}(q)\,e^{i q\cdot y}\, G_{1,2}^{2,0}\left(\frac{k_{\gamma}\,z^2}{2}\left|
\begin{array}{c}
 \frac{q^2}{2 k_{\gamma}}+1 \\
 0,1 \\
\end{array}
\right.\right)-\frac{1}{2} C^2_{\mu}(q)\,e^{i q\cdot y}\, k_{\gamma}\, z^2 \, _1F_1\left(1-\frac{q^2}{2 k_{\gamma}};\,2;\,-\frac{k_{\gamma}\,z^2}{2}\right)\,,
\end{eqnarray}
\noindent where 
\begin{equation}
    G_{p,q}^{m,n}\left(z\left|
\begin{array}{c}
 a_1 \cdots a_p \\
 b_1 \cdots b_q \\
\end{array}
\right.\right)\,\,\, {\rm and}\,\,\, _1F_1 (a; b; z)\nonumber 
\end{equation}

\noindent are the the Meijer G function and the Kummer confluent hypergeometric function, respectively. By imposing the boundary condition $\left.\phi_{\mu}(z,q)\right|_{z=0} = \eta_{\mu} e^{i q\cdot y}$, that implies $C_\mu^2(q)=0$, and considering normalizable (square integrable) solutions, one can write:
\begin{eqnarray}\label{phimunorm}
\phi_\mu(z,q)&=&-\frac{\eta_{\mu} e^{i q\cdot y}}{2} \, k_{\gamma}\, z^2 \, \Gamma{\left[1 - \frac{q^2}{2k_{\gamma}}\right]}\; {\cal U} \left(1-\frac{q^2}{2 k_{\gamma}};\,2;\,-\frac{k_{\gamma}\,z^2}{2}\right)\nonumber \\
&\equiv& -\frac{\eta_{\mu} e^{i q\cdot y}}{2}\, \mathcal{B}(z,q)\,, 
\end{eqnarray}
\noindent where $\Gamma[a]$ is the Gamma function and ${\cal U} (a, b,z)$ is the Tricomi hypergeometric function  \cite{abramowitz}. 
This equation represents the solution for the electromagnetic field that will be used to compute the pion form factor.

\section{Pion form factor}\label{pion}

The pion form factor is one of the most valuable QCD quantities related to the transition from the non-perturbative to the perturbative regime, appearing at large transferred momentum, $q$. In the electromagnetic case, the pion form factor comes from the annihilation or scattering of leptons interacting with charged pions. Specifically, it is defined from the photon-charged pions three-body vertex. Supposing a lepton scattering, we can write the corresponding amplitude as 
\begin{equation}
\mathcal{M}=\frac{1}{q^2}\,i\,Q\,\bar{u}(k_2)\,\gamma_\mu\,u(k_1)\,\langle \pi^{\pm}(p_2)\left|J_\pi^\mu(0)\right|\pi^{\pm}(p_1)\rangle,    
\end{equation}
\noindent where $Q$ stands for the lepton electric charge, the four-vectors $k_i$ and $p_i$ labels the leptons and pions momenta, $q=p_2-p_1$ is the virtual photon momentum which is the momentum transfer of the process, and $J_\pi^\mu$ is pion EM current.   The matrix element $\langle \pi^{\pm}(p_2)\left|J_\pi^\mu(0)\right|\pi^{\pm}(p_1)\rangle$ describes the pion-photon vertex, and it has a general Lorentz  structure defined in terms of the pions momenta as follows

\begin{equation}\label{pion-FF-M}
 \langle \pi^{\pm}(p_2)\left|J_\pi^\mu(0)\right|\pi^{\pm}(p_1)\rangle=c_{\pi^\pm}\,(p_1+p_2)^\mu\,F_\pi(q^2).   
\end{equation}
\noindent with $c_{\pi^\pm}$ the vertex coupling constant that normalizes the pion form factor $F_\pi (q^2)$. Notice this structure ensures that gauge, time reversal, parity, and Lorentz invariance is fulfilled. 

\begin{figure}[!ht]
  \centering
  \includegraphics[scale = 0.35]{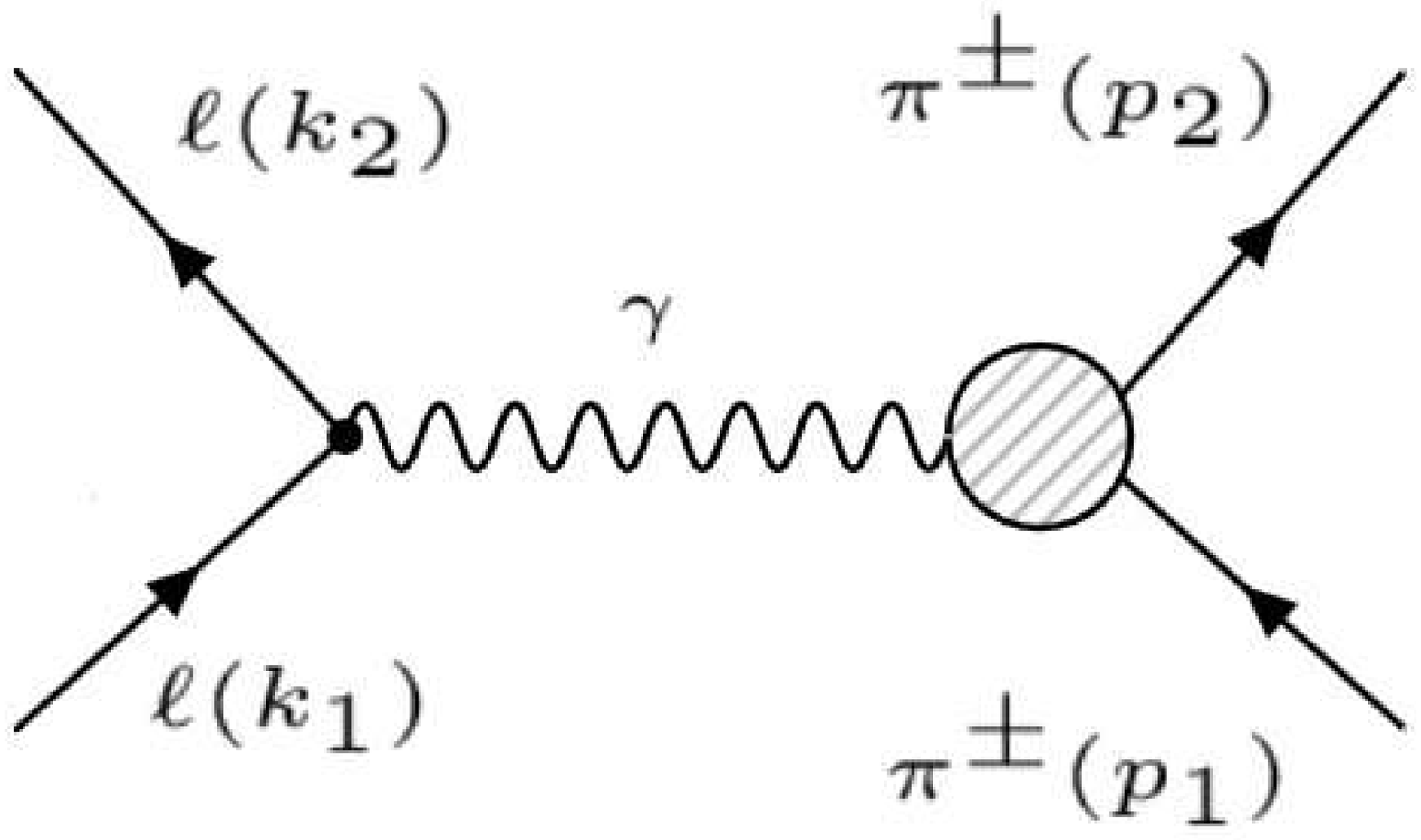} 
\caption{Feynman diagram representing the scattering pions and leptons via the exchange of a virtual photon. The shaded blob represents the effective vertex used to define the electromagnetic pion form factor.}
\label{dis}
\end{figure}

The Fig. \ref{dis} represents the Feynman diagram of a scattering between a pion and a lepton through the exchange of a virtual photon.

At the holographic level, we will suppose this pionic matrix element lives at the conformal boundary; thus, we will follow the AdS/CFT standard technics to write down an expression for the pion form factor using the deformed geometric model described above.

Let us focus on the holographic calculation:  the three-point effective vertex in the bulk, dual to the two-pion-photon vertex, is defined \emph{via} the following structure when the minimal coupling is imposed \cite{Polchinski:2001ju}:

\begin{equation}\label{eff-action}
S_\text{eff}=g_\text{eff}\int{d^5x\,\sqrt{-g_{\pi}}\,g^{mn}_{\pi}\,\phi_m(x,z)\,\left[X_{p_1}(x,z)\,\partial_m\,X^*_{p_2}(x,z)-X^*_{p_2}(x,z)\,\partial_m\,X_{p_1}(x,z)\right]},    
\end{equation}
\noindent where the bulk fields $X(x,z)$ and $\phi_m(x,z)$, dual to the pion and the virtual photon, are defined in terms of the Eqns. \eqref{eom_esc_2} and \eqref{solem}. The coupling $g_\text{eff}$ is a constant that fixes units in the effective action. These bulk fields can be spanned in terms of  waves in AdS as 
\begin{eqnarray*}
X(x,z)&=&e^{-i\,p\cdot x}\,v(z),\\
\phi_\mu(x,z)&=&\eta_\mu\,e^{-i\,q\cdot x}\,\mathcal{B}(z).
\end{eqnarray*}

Putting these definitions into the effective action and after some calculations, we will obtain 
\begin{equation}
S_\text{eff}=i\,(2\,\pi)^4\, \delta^4(q-(p_2-p_1))\,\eta^{\mu\nu}\eta_\mu\,\left(p_1+p_2\right)_\nu\,g_\text{eff}\,\int{dz\,e^{3\,A_{\pi}(z)}\,v(z)\,\mathcal{B}(z,q)\,v(z)},   
\end{equation}
\noindent where the delta appears due to the four-momentum conservation at the vertex, and $A(z)$ is the warp factor defined in Eq. \eqref{A}.

Finally, from the effective action, isolating all of the terms associated with momentum conservation, we can extract the pion matrix element defined in Eq. \eqref{pion-FF-M}, allowing us to write the pion form factor as 

\begin{equation} \label{form-fact-1}
  F_\pi(q^2) = \int{dz\,e^{3\,A_{\pi}(z)}\,v(z)\,\mathcal{B}(z,q^2)\,v(z)}. 
\end{equation}

\noindent with $v(z)$ is the scalar normalizable mode dual to pion. The field $\mathcal{B}(z,q)$ is the vector non-normalizable mode (bulk-to-boundary propagator) dual to the virtual photon.

Notice that under the boundary conditions, we expect that $\phi_m(x,z\to 0)=\eta_m\,e^{-i\,q\cdot x}$, yielding $\mathcal{B}(z\to 0,q^2)=\mathcal{B}(z, q^2\to 0)=1$. These conditions imply naturally that $F(q^2\to 0)=1$, since $v(z)$ is normalized. Therefore, we conclude that $c_{\pi^\pm}=g_\text{eff}=1$ in our approach.   

We can go further by introducing the Schrodinger-like modes for the bulk scalar field, i.e., $v(z)=e^{-3\,A_{\pi}(z)/2}\,\psi(z)$. Since we consider that the charged pion at the boundary is dual to the scalar field ground state, we will restrict ourselves to consider only $\psi_1(z)$ (See figure \ref{fig:one}). Thus, the final expression for the form factor in our case is

\begin{equation}\label{form-fact-2}
F_\pi(q^2)=\int{dz\,\psi_1(z)\,\mathcal{B}(z,q^2)\,\psi_1(z)}.     
\end{equation}

This integral can not be computed by analytical approaches and then we will solve it numerically. After computing the pion form factor in Eq. \eqref{form-fact-2}, one can get the pion radius which is given by: 
\begin{equation}\label{raio}
    \langle r^2_{\pi} \rangle = - 6 \,\left.\frac{d F_{\pi}(q^2)}{d q^2}\right|_{q^2=0}\,.
\end{equation}

In the next section we will present our results in comparison with the experimental data and some theoretical works (holographic and nonholographic).


\section{Numerical results for the Pion form factor}\label{numerical}

In this section we will explore and comment on our numerical results obtained from our deformed AdS/QCD model and compare them with experimental data as well as some theoretical approaches.

Just before to present our results it is worthwhile to make some comments on previous results for pion form factor which will guide us in order to explain our results. Let us start our discussion by experimental data on pion form factor. As one can see in Refs. \cite{Ackermann:1977rp, Bebek:1977pe, Brauel:1979zk, Amendolia:1986wj, Horn:2006tm, Tadevosyan:2007yd} there is  vast collection of those data obtained by renowned  collaborations in the last 40 years. However one should notice that mostly of these data are related to soft processes (low $q^2$), or intermediate ones,  and the few data related to hard processes (large $q^2$), so far, are not reliable.

In the framework of theoretical works, we will start with the iconic papers in Refs. \cite{Brodsky:1973kr, Matveev:1973ra}, Brodsky-Farrar and Matveev-Muradian-Tavkhelidze, respectively, predicted a scaling law at large transverse momentum meaning for large $q^2$ regime, pion form factor should behave as  $F_{\pi}(q^2 \to \infty) \sim {q^{-2}}$. Another important prediction related to the pion form factor at at moderately large $q^2$ can be seen in Ref. \cite{Efremov:2009dx} where Efremov and Radyushkin, taking into account quark counting rule (QCR) argue the the pion form factor behaves as $F_{\pi}(q^2 \to \infty) \sim (q^{-2})^{n-1}$, where $n$ corresponds the number of the valence quarks in a composed system. Also in Ref. \cite{Efremov:2009dx} the authors mention that in this moderately large $q^2$ region the contribution coming from
Feynman mechanism is damped by the Sudakov form factor meaning an abrupt decreasing of the pion form factor while $q^2$ increasing. For large $q^2$  regimes the process recovers  $F_{\pi}(q^2 \to \infty) \sim {q^{-2}}$ behaviour. It is worthwhile to remember that other important theoretical works were cited in Sec.  \ref{intro}. In particular, for the comparison with the results achieved in this work, we will focus in approaches showed in Refs. \cite{Nesterenko:1982gc, Geshkenbein:1998gu, Maris:2000sk, Bakulev:2004cu, Choi:2006ha, Lomon:2016eyp}.
 
Studies related to the pion form factor within AdS/QCD program were motioned in Sec. \ref{intro}. For sake of completeness we will present them again here (see Refs. \cite{Grigoryan:2008up, Brodsky:2007hb, Kwee:2007dd, Vega:2008te, Bayona:2010bg, Zuo:2009hz, Stoffers:2011xe, Gutsche:2014zua}). Such works achieved a good agreement with others found in the literature specially for soft or intermediate processes. \textcolor{black}{In particular, the Refs. \cite{Grigoryan:2008up, Brodsky:2007hb} addressed two important questions related to bottom-up models, as the softwall one. The first question was presented in Ref. \cite{Grigoryan:2008up} where the authors have noticed that if one considers the conformal dimension as $\Delta=3$, the behaviour for the corresponding form factors for vector mesons scales as $1/q^4$, which is inconsistent with the QCD sum rules. The second question was brought in Ref. \cite{Brodsky:2007hb}, where the authors  proposed that the conformal dimension associated with the bulk scalar field should be reinterpreted as the twist, i.e., the number of hadronic constituents, to get the correct results. Such a  reinterpretation implies that the near-to-the-boundary behavior of the bulk modes also changes.}

\textcolor{black}{The motivation for our work can be enclosed in a single question:} \textcolor{black}{why the conformal dimension used in holography at constant time is interpreted as scaling dimension, while when we deal with form factors in light-front holography, it should be reinterpreted as the number of constituents?}. \textcolor{black}{In order to answer this question, our work explores another possibility: instead of reformulating the meaning of the conformal dimension, we modify the energy scale associated with the virtual photon in the scattering process. Recall that the photon energy scale is not related with the pion confinement process, since both particles are described by two different geometries. The photon scale measures the momentum transfer between the virtual photon and the pion at the interaction vertex. Thus, we can expect this scale to be a function of the transferred momentum, $q$. }

\textcolor{black}{After this brief digression we are able to present our numerical results within our deformed AdS/QCD model and compare it with the available experimental data, as well as other non-holographic and holographic results. In Section \ref{d3} and  \ref{qk} we will present our results for both pion form factor and pion radius considering the canonical conformal dimension as $\Delta =3$. In Section \ref{d3}, we take $k_\gamma$ independent of $q$,  and in Section \ref{qk}, the energy scale $k_\gamma$ associated with the virtual photon in the scattering process is assumed to be a function of $q$. In particular, the results of section \ref{qk} for pion form factor and pion radius are in agreement with those found in the literature.}

\subsection{Pion form factor and pion radius for $\Delta=3$}\label{d3}

In this section we will present the results achieved within our deformed model considering the scaling dimension $\Delta =3$ for the operators which will represent the pion at the boundary.

\begin{center}
\begin{figure*}
  \begin{tabular}{c c}
  \includegraphics[scale = 0.25]{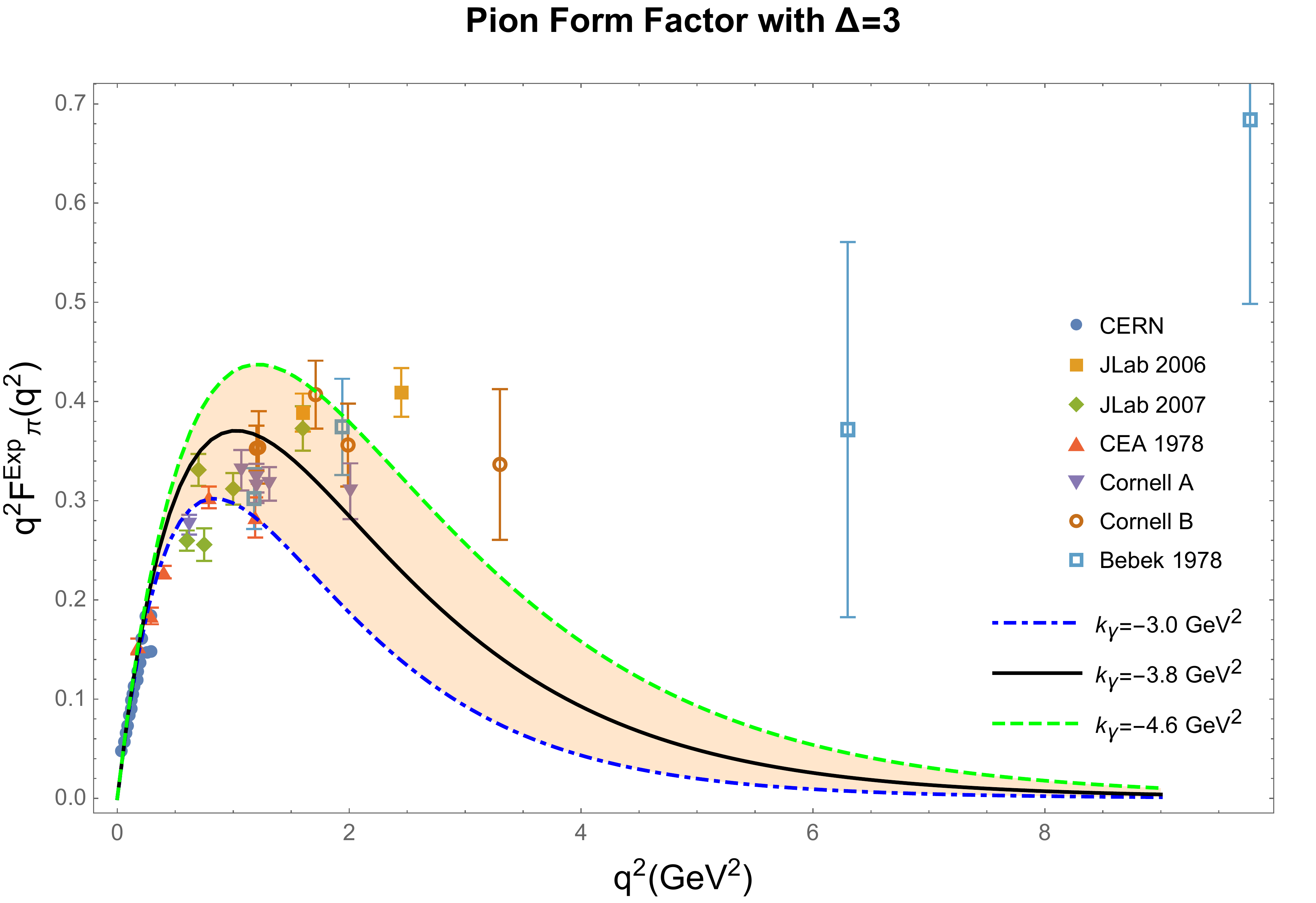} \\
    \includegraphics[width=3.1 in]{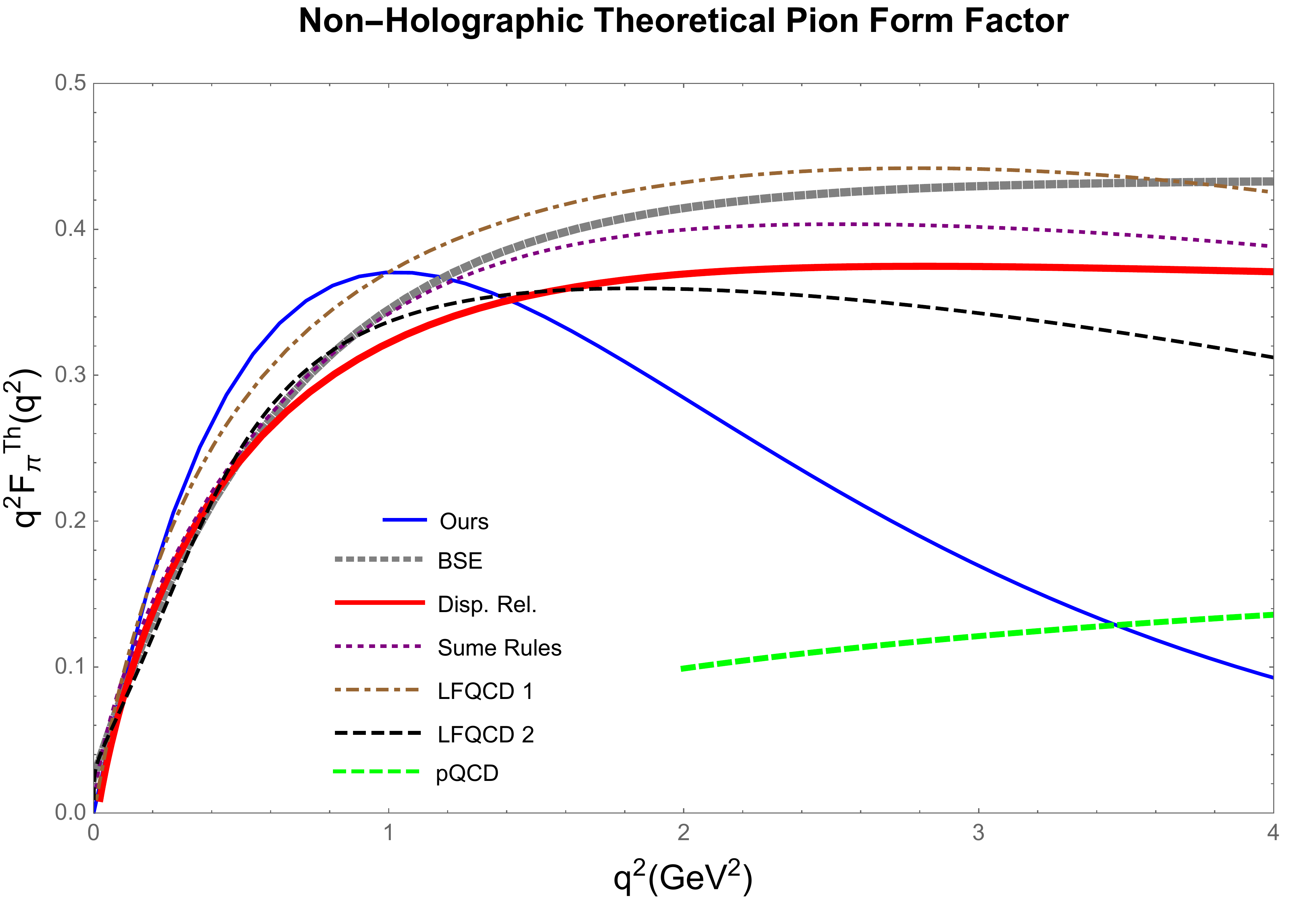}
      \includegraphics[width=3.1 in]{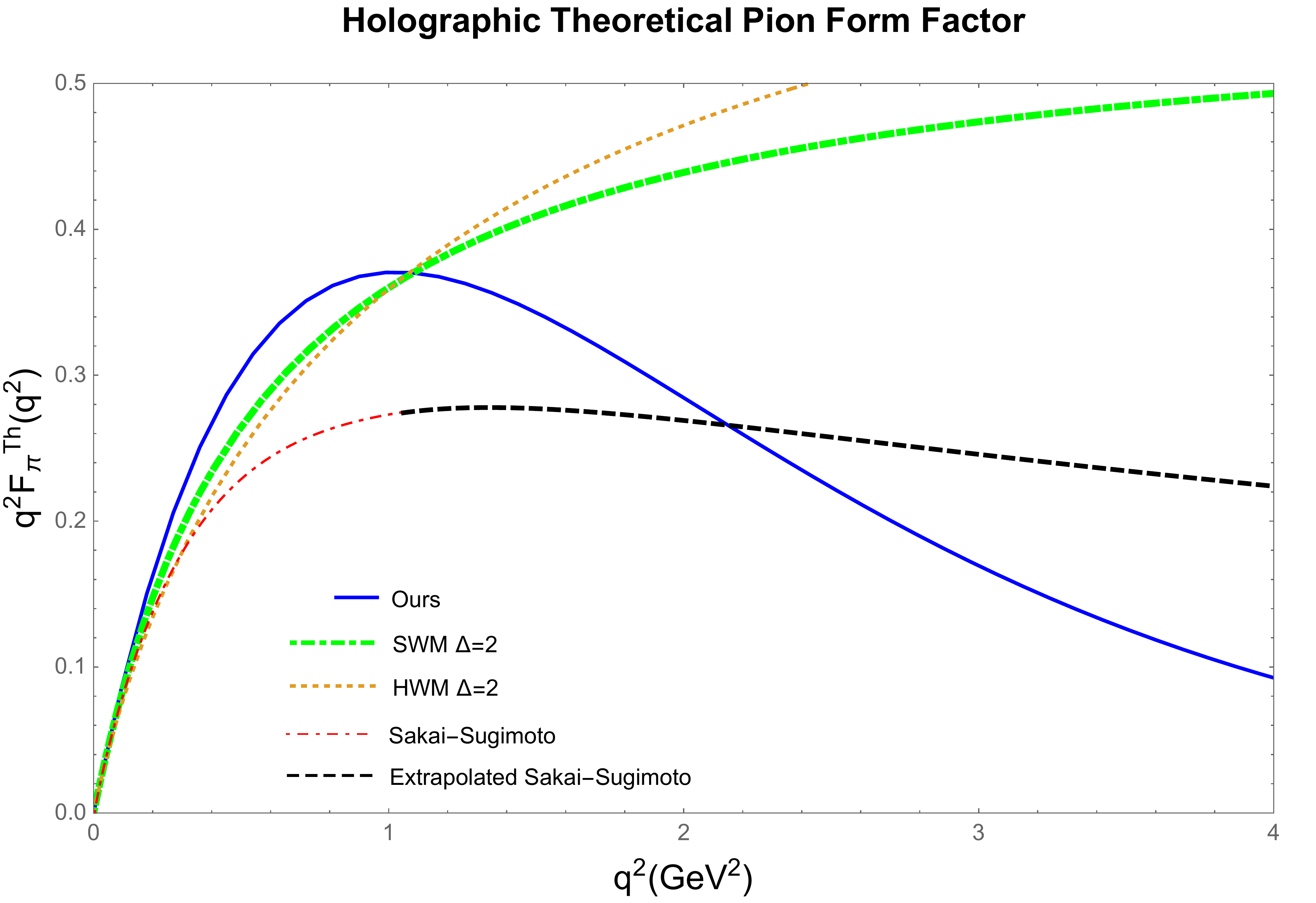}\\
  \end{tabular}
\caption{The upper panel compares our results for the pion form factor with the available experimental data \cite{Ackermann:1977rp,Bebek:1977pe,Amendolia:1986wj,Brauel:1979zk,Horn:2006tm,Tadevosyan:2007yd}. In the lower panels we have a comparison of our results with non-holographic models (left panel) such as BSE \cite{Maris:2000sk}, perturbative QCD \cite{Bakulev:2004cu}, dispersion relations \cite{Geshkenbein:1998gu}, sum rules \cite{Nesterenko:1982gc}, and LFQCD \cite{Choi:2006ha}. In the lower right panel, we depict a comparison with other holographic models such as hardwall and softwall with $\Delta=2$ \cite{Brodsky:2007hb}, and Sakai-Sugimoto/extrapolated Sakai-Sugimoto \cite{Bayona:2010bg}. In our results we have taken $k_\gamma=-3.8$ GeV$^2$.}
\label{fig:two}
\end{figure*}
\end{center}

Computing numerically the integral in Eq. \eqref{form-fact-2}, one get the pion form factor, as a function of the squared four-momentum, in comparison with available experimental data, presented in the upper panel of Fig. \ref{fig:two}.
%
%

Instead to fix only one value for the parameter $k$ and gets a single curve, we rather choose to present not only one curve, but a region (colored). Inside this region, our model works in agreement with the experimental data. In particular, this agreement is for small $q^2$.

In the lower panels in Fig. \ref{fig:two} we compare our results with other theoretical non-holographic approaches found in the literature (left panel), as well as with other holographic ones (right panel).
It is worthwhile to mention that the pion radius computed by using our model, from Eq. \eqref{raio}, considering $\Delta=3$, is $r_\pi=0.458$ fm with an error around 30$\%$ compared to the experimental one \cite{pdg}. Besides one can also notice that our model with this consideration does not seem to capture the predicted scaling law $F_{\pi}(q^2 \to \infty) \sim {q^{-2}}$. In the appendix \ref{app-2} we provide a brief review on this scaling law within softwall context.

In the next section we will propose a modification in our model, different from the one proposed by Brodsky and Teramond in Ref. \cite{Brodsky:2007hb}, in order to accommodate the experimental/theoretical results for the pion radius and the scaling law  $F_{\pi}(q^2 \to \infty) \sim {q^{-2}}$.

\subsection{Pion form factor and pion radius for $\Delta=3$ and $k$ dependent of the momentum}\label{qk}

In ref. \cite{Brodsky:2007hb},  the authors are using the light-front approach, based on in the softwall model with the dilaton field $\Phi(z)=\kappa^2\,z^2$, which obeys to the AdS/CFT dictionary, where the conformal dimension of the operator which represents the scalar field should be $\Delta=3$. However, their solution was based on considering their conformal dimension as the twist dimension $(\Delta = \tau = 2)$, for the scalar particles which have spin $J=0$. The twist is related to the number of the constituents of the hadron. By doing this they achieved the correct scaling law as can seen through Eq. \ref{scaling} in appendix \ref{app-2}.

\begin{center}
\begin{figure*}
  \begin{tabular}{c c}
  \includegraphics[scale = 0.25]{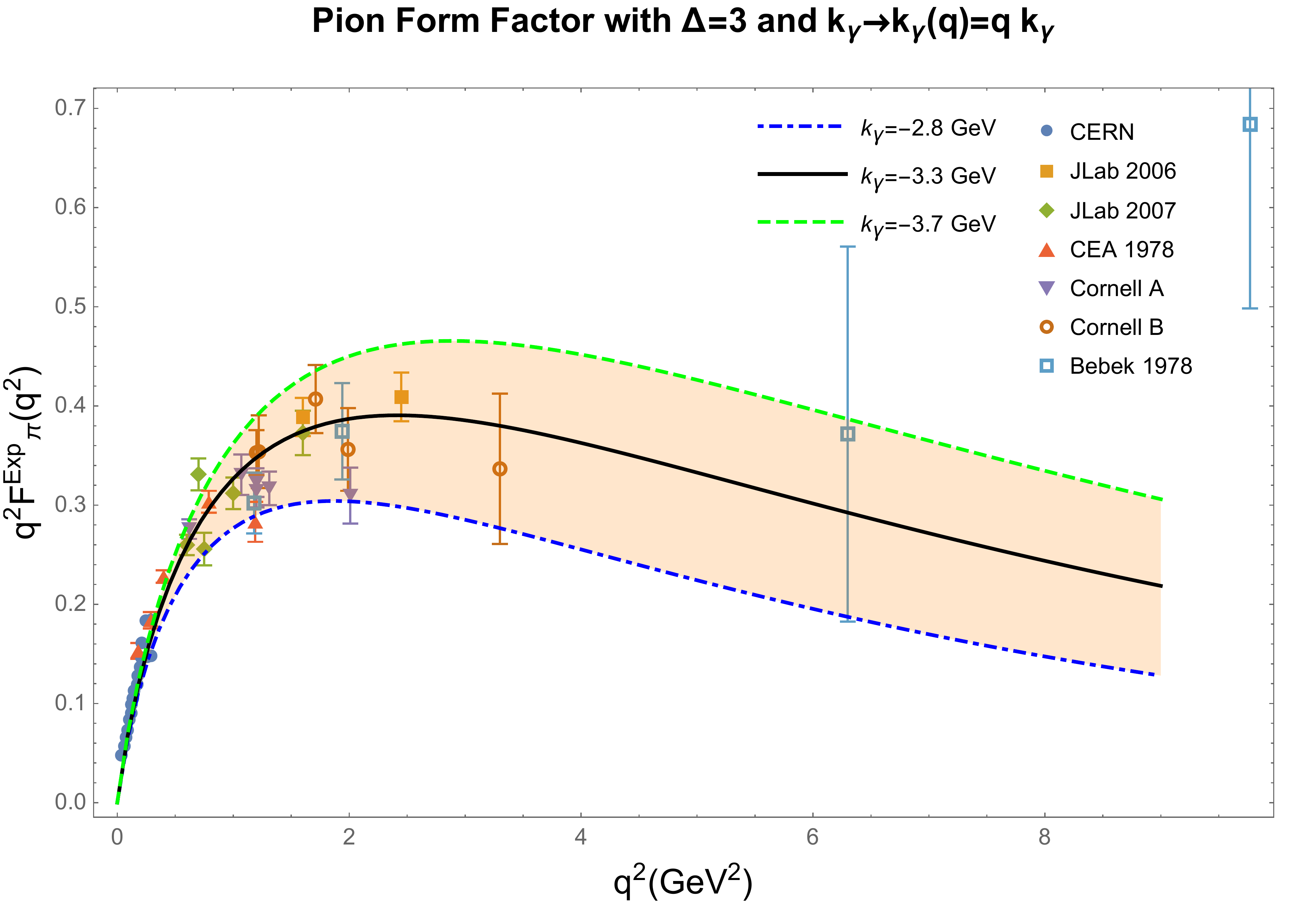} \\
    \includegraphics[width=3.1 in]{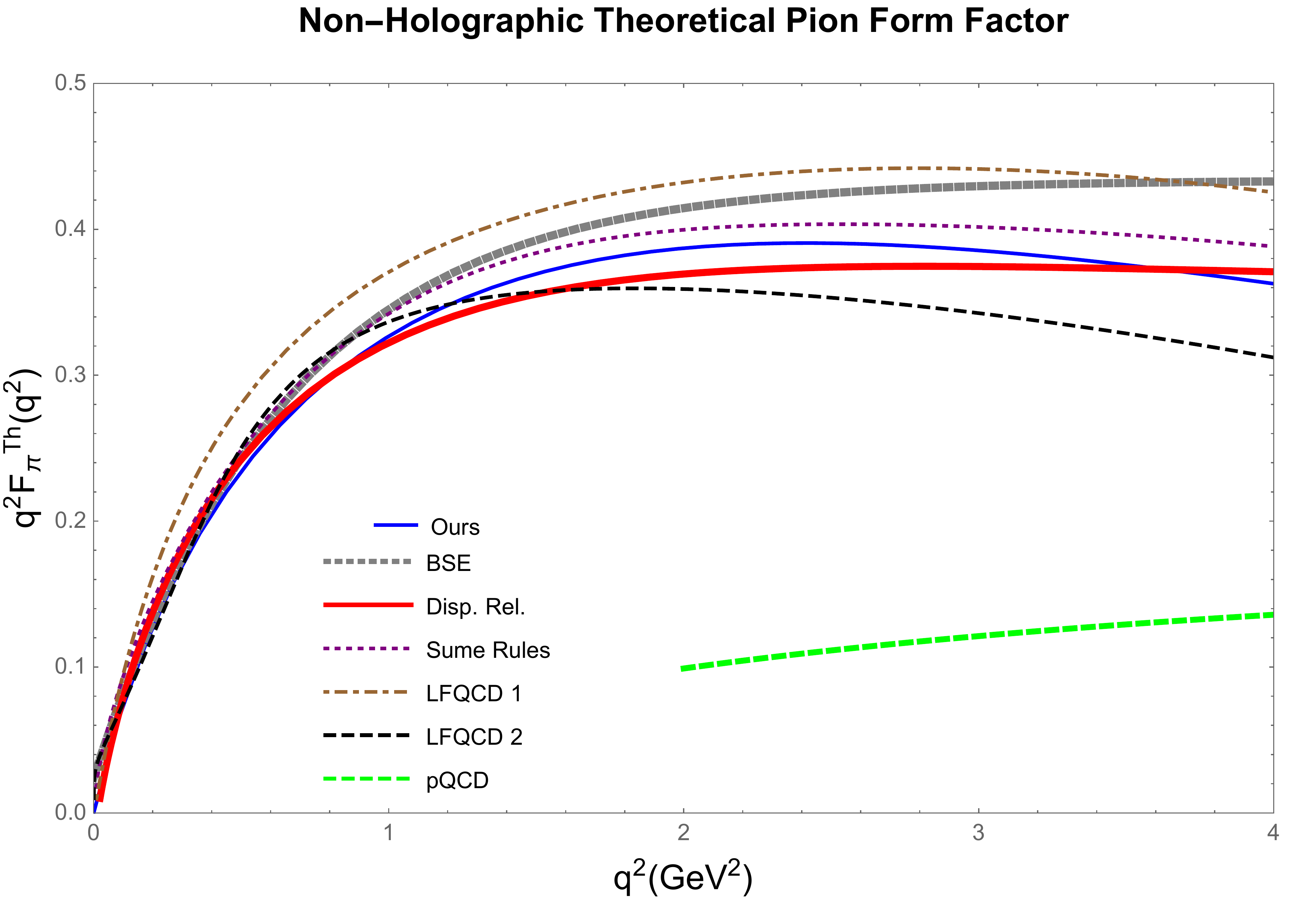}
      \includegraphics[width=3.1 in]{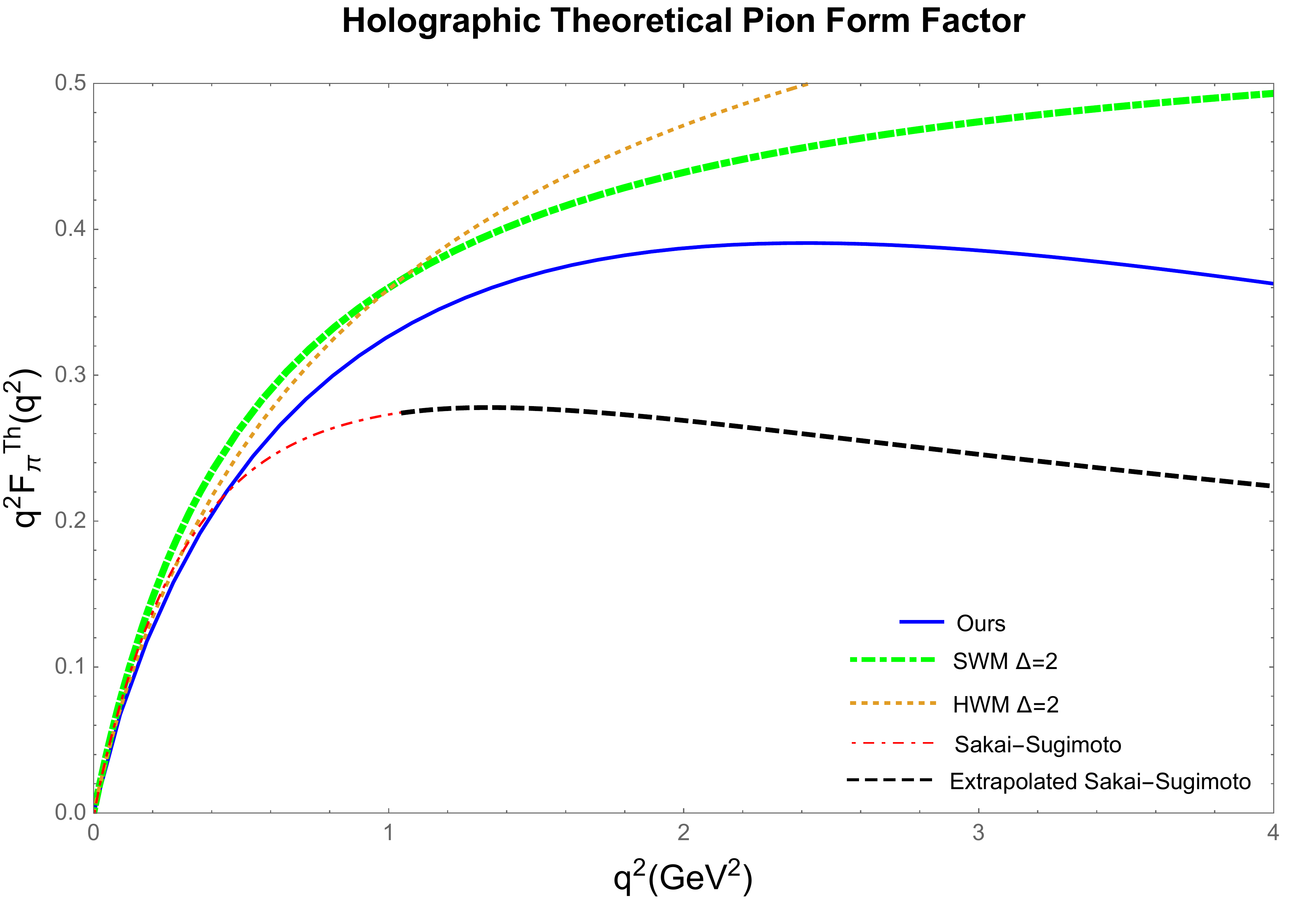}\\
  \end{tabular}
\caption{The upper panel compares our results for the pion form factor with the available experimental data \cite{Ackermann:1977rp,Bebek:1977pe,Amendolia:1986wj,Brauel:1979zk,Horn:2006tm,Tadevosyan:2007yd} using the proposed scaling for $k_\gamma$. In the lower panels we depict a comparison of our results with non-holographic models (left panel) such as BSE \cite{Maris:2000sk}, perturbative QCD \cite{Bakulev:2004cu}, dispersion relations \cite{Geshkenbein:1998gu}, sum rules \cite{Nesterenko:1982gc}, and LFQCD \cite{Choi:2006ha}. In the lower right panel, we show a comparison with other holographic models such as hardwall and softwall with $\Delta=2$ \cite{Brodsky:2007hb}, and Sakai-Sugimoto/extrapolated Sakai-Sugimoto \cite{Bayona:2010bg}. In our results, for lower panels we have taken $k_\gamma=-2.8$ GeV$^2$.}
\label{fig:three}
\end{figure*}
\end{center}

Here, in this work,  we will not resort to the twist dimension. This means that we will consider $\Delta=3$. We will propose  that our free parameter could depends on the scale of energy in the scattering process with profile $k_\gamma \to k_\gamma(q)=q\,k_\gamma$.

Assuming this simple profile we could accommodate experimental/theoretical results for small, intermediate and large $q^2$. Besides, the value we have found for the pion radius, $r_{\pi} = 0.671$ fm, is in agreement with the experimental data \cite{pdg} presenting a relative error of 2.0 \%,   and the scaling law  $F_{\pi}(q^2 \to \infty) \sim {q^{-2}}$ is recovered. These results can be seen in Fig. \ref{fig:three}.

The motivation behind this particular choice of $k_\gamma$ comes from the analysis of the pion form factor $F_\pi(q)$ in the softwall model context, with quadratic dilaton $\Phi(z)=\left|k_\gamma\right|\,z^2$, see appendix \ref{app-1}. Then, eq. \eqref{FFF}  has the following form: %

\begin{equation}
F_\pi(q)=\frac{32\,k_\gamma^2}{\left(q^2+4\,\left|k_\gamma\right|\right)\left(q^2+8\,\left|k_\gamma\right|\right)},  
\end{equation}
\noindent where $\left|k_\gamma\right|$ is our deformation slope in energy squared units. 

In general, this behavior is not restricted to the softwall model only. It is expected to appear in any other AdS/QCD model that has quadratic static dilaton or geometric deformation.  Therefore we propose a rescaling of our photon slope as $k_\gamma\to k_\gamma(q)=q \,k_\gamma$, that recovers the expected phenomenological behavior $\left.F_\pi\right|_{q^2\to\infty}\to q^{-2}$, produces good agreement with the experimental/theoretical data and gives a very low relative error for the pion radius.  

\section{Conclusions} \label{conc}
In this work we have discussed the pion form factor calculation in the context of a geometric deformed AdS/QCD model, considering the conformal dimension $\Delta=3$ associated with the operator that creates pions in the boundary.

In this formalism with fixed $k_\gamma$, the scaling law for the pion form factor at $q^2\to \infty$ is not captured properly. Despite that numerical results are reasonable for the intermediate $q^2$ region, as Figure \ref{fig:two} shows, the pion radius is also not well fitted. 

In order to improve our results, we still consider $\Delta=3$, and propose a rescaling in the parameter $k_\gamma$ with the transferred momentum $q$, i.e., $k_\gamma\to k_\gamma(q)=q\,k_\gamma$ that fixes the form factor behavior at large $q^2$. 

This assumption seems to be natural since, as we discussed in appendix \ref{app-2}, large $q^2$ behavior comes from the analysis of the pion eigenmodes near the conformal boundary, where the dilaton or the deformation do not have any effect.

The approach exposed here is another possibility rather than the proposal presented by Brodsky and de Teramond \cite{Brodsky:2007hb}, where they consider $\Delta=2$ in light-front softwall model. 

It is worth to mention that our approach seems to present a Sudakov suppression similar to the one  pointed out in Ref. \cite{Efremov:2009dx}, which is not manifest in other bottom-up AdS/QCD models, as the hardwall and the softwall \cite{Brodsky:2007hb}. 




\begin{acknowledgments}

A. V. and  M. A. M. C.  would like to thank the financial support given by FONDECYT (Chile) under Grants No. 1180753  and No. 3180592,  respectively. D.L. is supported by the National Natural Science Foundation of China (11805084), the PhD Start-up Fund of Natural Science Foundation of Guangdong Province (2018030310457) and Guangdong Pearl River Talents Plan (2017GC010480). H.B.-F. is partially supported by Coordenação de Aperfeiçoamento de Pessoal de Nível Superior (CAPES) under finance code 001,   and Conselho Nacional de Desenvolvimento Científico e Tecnológico (CNPq) under Grant No. 311079/2019-9.

\end{acknowledgments}

\appendix

\section{Large $q^2$ analysis in the AdS deformed background}\label{app-2}

Let us consider the photon bulk to boundary propagator given by eqn. \eqref{phimunorm}. Following \cite{Brodsky:2007hb}, we found an integral representation for the Tricomi function 

\begin{equation}
\Gamma(a)\,\mathcal{U}(1,b,z)=\int_0^\infty{dt\,e^{-z\,t}\,t^{a-1}\,\left(1+t\right)^{b-a-1}},
\end{equation}

\noindent which in our particular case reads as

\begin{equation}
\Gamma\left(1+\frac{q^2}{2\,|k|}\right)\,\mathcal{U}(1+\frac{q^2}{2\,|k|},2,\frac{|k|z^2}{2})= \int_0^\infty{dt\,e^{-\frac{|k|\,z^2}{2}}\left(\frac{1+t}{t}\right)^{-\frac{q^2}{2\,|k|}}},   
\end{equation}

\noindent where we have redefined $k\equiv -|k|$, since it is defined as a negative quantity \emph{per se}. Therefore, the progatator $\mathcal{B}(z,q^2)$ is written as
 
 \begin{equation}
\mathcal{B}(z,q^2)=\frac{|k|\,z^2}{2}\,\int_0^\infty{dt\,e^{-\frac{|k|\,z^2}{2}}\,\left(\frac{1+t}{t}\right)^{-\frac{q^2}{2\,|k|}}}.     
 \end{equation}
 
In order to consider large $q^2$ limit, we need to do the transformation $t=\frac{q^2}{2\,|k|}\,\mu$ in the integral representation defined above (See Appendix C in Ref. \cite{Brodsky:2007hb}), obtaining 

\begin{equation}
\mathcal{B}(z,q^2)=\frac{|k|z^2}{4}\,\int_0^\infty{d\mu\,e^{-\frac{q^2\,z^2}{4}\,\mu}\,\left(\frac{1+\frac{q^2}{2|k|}\,\mu}{\frac{q^2}{2\,|k|}}\right)^{-\frac{q^2}{2\,|k|}}}.    
\end{equation}

Now, we can perform the large $q^2$ limit, yielding 

\begin{equation}
\mathcal{B}(z,q^2\to \infty)=\frac{q^2\,z^2}{4}\int_0^\infty{d\mu\,e^{-\frac{q^2\,z^2}{4}\mu}\,e^{-\frac{1}{\mu}}}.
\end{equation}

On the other hand, we have the following integral representation for Bessel functions 

\begin{equation}
K_\nu(z)=\frac{1}{2} \left(\frac{z}{2}\right)^\nu\int_0^\infty{dx\,\frac{e^{-x-\frac{z^2}{4\,x}}}{t^{\nu+1}}}.   
\end{equation}

 If we fix $\nu=-1$ we arrive to
 
\begin{equation}
\mathcal{B}(z,q^2\to \infty)=q\,z\,K_1(q\,z),    
\end{equation}
 
\noindent since parity of Bessel functions: $K_{-\nu}(z)=K_\nu(z)$. From holographic grounds, taking $q^2\to \infty$ implies $z\to0$, since we have $z\to 1/q$. In this limit, the form factor is written as 

\begin{eqnarray}\notag
\left.F_\pi(q^2)\right|_{q^2\to \infty}&=& \int_{z\to1/q}{\frac{dz}{z^3}\,\left(z^{\Delta}\right)^2\,q\,z\,K_1(q\,z)}\\ \notag
&=& \int_{z\to1/q}{\frac{dz}{z^3}\,\left(z^{\Delta}\right)^2\,\left[1+\frac{q^2}{4}\left(-1+2\,\gamma_e+2\log\frac{q\,z}{2}\right)\,z^2+\mathcal{O}\left(z^4\right)\right]}\cr   
&=&\frac{1}{8\,\Delta^2\left(\Delta-1\right)}\left(\frac{1}{q^2}\right)^{\Delta-1}\left[1+\gamma_e\left(\Delta-1\right)\Delta-\Delta^2\left(\log 4-3\right)+\Delta\,\log 4\right]\\
\end{eqnarray}

\noindent where we have taken the low $z$-limit of the pion eigenmodes and the warp factor, and also the power series of the Bessel function $K_\nu(z)$. Finally, that the form factor in this case scales as 
 
\begin{equation}\label{scaling}
\left.F_\pi(q^2)\right|_{q\to\infty}\to \left(\frac{1}{q^2}\right)^{\Delta-1}.    
\end{equation} 
 
Notice this scaling is expected from the solutions obtained for the $\Delta=3$ within the SWM.  Recall the dilaton or the deformation do not affect the low-$z$ behavior of the bulk eigenmodes.

\section{Pion form factor in the original softwall model}\label{app-1}

Let us consider the softwall model (SWM) \cite{Karch:2006pv}, with $\Phi(z)=\kappa^2\,z^2$, applied to scalar mesons in order to compute the pion form factor, following the same idea proposed in Ref. \cite{Brodsky:2007hb}, but without considering the twist dimension. In general, for the scalar mesons, the scaling dimension $\Delta=3$ fixes the bulk mass to be $M_5^2\,R^2=-3$. In such a case, solutions are written as associated Laguerre polynomials \cite{Colangelo:2008us}. In this scenario, the pion is associated with the lowest eigenmode, which is $n=0$, giving a cubic polynomial in $z$, i.e., 

\begin{equation}
    \psi_0(z)=\sqrt{\frac{2\,\kappa^4}{R^3}}\,z^3,
\end{equation}

\noindent where $\kappa$ stands for the dilaton slope that defines the scalar meson linear Regge trajectory and $R$ is the AdS curvature radius. The bulk-to-boundary propagator $\mathcal{V}(z,q)$ associated with the photon in the softwall model (SWM obeys the following equation of motion

\begin{equation}
\partial_z\left[\frac{e^{-\kappa^2\,z^2}}{z}\,\mathcal{V}'(z)\right]+(-q^2)\frac{e^{-\kappa^2\,z^2}}{z}\,\mathcal{V}(z)=0.   
\end{equation}

The solution of this equation, in general, is written in terms of Tricomi hyperconfluent functions as

\begin{equation}
\mathcal{V}(z,q)=\Gamma\left(1+\frac{q^2}{4\,\kappa^2}\right)\,U\left(\frac{q^2}{4\,\kappa^2},0,\kappa^2\,z^2\right)=\kappa^2\,z^2\,\int_0^1{\frac{dx}{(1-x)^2}\,e^{-\frac{\kappa^2\,z^2\,x}{1-x}}\,x^{\frac{q^2}{4\,\kappa^2}}},    
\end{equation}

\noindent where we have used the integral representation for the bulk-to-boundary propagator, suggested by Grigoryan and Radyushkin, in Ref. \cite{Grigoryan:2007my}, in the right part of the equation. From this integral definition for the pion form factor, we have

\begin{eqnarray}
F_{\pi}(q^2)&=&R^3\,\int_0^\infty{dz\,\frac{e^{-\kappa^2\,z^2}}{z^3}\,\psi_0^*(z)\,\mathcal{V}(z,q)\,\psi_0(z)}\\
&=&2\,\kappa^6\,\int_0^1{\frac{dx}{(1-x)^2}\,x^{\frac{q^2}{4\,\kappa^2}}\,\int_0^\infty{dz\,z^5\,e^{-\frac{\kappa^2\,z^2}{1-x}}}}\\
&=&\frac{32\,\kappa^4}{\left(q^2+4\,\kappa^2\right)\left(q^2+8\,\kappa^2\right)}\label{FFF}
\end{eqnarray}

As we are considering from the beginning of this section that the conformal dimension $\Delta=3$, one should notice that the form factor in this scenario suggests a $q^{-4}$ behavior which is not expected from the known particle phenomenology \cite{Brodsky:1973kr}. Moreover, since any deformation or dilaton used in the AdS background does not modify the UV behavior of the bulk solutions, we can say that this is a general feature of the softwall-like AdS/QCD models. 

However if we consider in Eq. \eqref{FFF}, $\kappa \to \sqrt{q}\, \kappa$, meaning the dilaton slope is now depending on $q$ or the energy scale in the process. And then, For large $q^2$, Eq.  \eqref{FFF}, behaves as:
\begin{equation}
    F_{\pi}(q^2) \sim \frac{1}{q^2}\,,
\end{equation}
fulfilling the expected scaling law even considering $\Delta =3$. Another possibility is to consider $\Delta=2$ as done in Ref. \cite{Brodsky:2007hb} in the context of Light-Front holography.


\begin{thebibliography}{99}
\bibitem{Amendolia:1986wj}
S.~R.~Amendolia \textit{et al.} [NA7],
``A Measurement of the Space - Like Pion Electromagnetic Form-Factor,''
Nucl. Phys. B \textbf{277}, 168 (1986)

\bibitem{Ackermann:1977rp}
H.~Ackermann, T.~Azemoon, W.~Gabriel, H.~D.~Mertiens, H.~D.~Reich, G.~Specht, F.~Janata and D.~Schmidt,
``Determination of the Longitudinal and the Transverse Part in pi+ Electroproduction,''
Nucl. Phys. B \textbf{137}, 294-300 (1978)
\bibitem{Bebek:1977pe}
C.~J.~Bebek, C.~N.~Brown, S.~D.~Holmes, R.~V.~Kline, F.~M.~Pipkin, S.~Raither, L.~K.~Sisterson, A.~Browman, K.~M.~Hanson and D.~Larson, \textit{et al.}
``Electroproduction of single pions at low epsilon and a measurement of the pion form-factor up to $q^2$ = 10-GeV$^2$,''
Phys. Rev. D \textbf{17}, 1693 (1978)
\bibitem{Tadevosyan:2007yd}
V.~Tadevosyan \textit{et al.} [Jefferson Lab F(pi)],
``Determination of the pion charge form-factor for Q**2 = 0.60-GeV**2 - 1.60-GeV**2,''
Phys. Rev. C \textbf{75}, 055205 (2007)
[arXiv:nucl-ex/0607007 [nucl-ex]].
\bibitem{Nesterenko:1982gc}
V.~A.~Nesterenko and A.~V.~Radyushkin,
``Sum Rules and Pion Form-Factor in QCD,''
Phys. Lett. B \textbf{115}, 410 (1982)
\bibitem{Geshkenbein:1998gu}
B.~V.~Geshkenbein,
``Pion electromagnetic form-factor in the space - like region and P phase delta(1) in one-dimension (s) of pi pi scattering from the value of the modulus of form-factor in the time - like region.,''
Phys. Rev. D \textbf{61}, 033009 (2000)
[arXiv:hep-ph/9806418 [hep-ph]].
\bibitem{Maris:2000sk}
P.~Maris and P.~C.~Tandy,
``The pi, K+, and K0 electromagnetic form-factors,''
Phys. Rev. C \textbf{62}, 055204 (2000)
[arXiv:nucl-th/0005015 [nucl-th]].
\bibitem{Bakulev:2004cu}
A.~P.~Bakulev, K.~Passek-Kumericki, W.~Schroers and N.~G.~Stefanis,
``Pion form-factor in QCD: From nonlocal condensates to NLO analytic perturbation theory,''
Phys. Rev. D \textbf{70}, 033014 (2004)
[erratum: Phys. Rev. D \textbf{70}, 079906 (2004)]
[arXiv:hep-ph/0405062 [hep-ph]].
\bibitem{Choi:2006ha}
H.~M.~Choi and C.~R.~Ji,
``Conformal symmetry and pion form-factor: Soft and hard contributions,''
Phys. Rev. D \textbf{74}, 093010 (2006)
[arXiv:hep-ph/0608148 [hep-ph]].
\bibitem{Lomon:2016eyp}
E.~L.~Lomon and S.~Pacetti,
``Analytic pion form factor,''
Phys. Rev. D \textbf{94}, no.5, 056002 (2016)
[arXiv:1603.09527 [hep-ph]].
\bibitem{Frezzotti:2008dr}
R.~Frezzotti \textit{et al.} [ETM],
``Electromagnetic form factor of the pion from twisted-mass lattice QCD at N(f) = 2,''
Phys. Rev. D \textbf{79}, 074506 (2009)
[arXiv:0812.4042 [hep-lat]].

\bibitem{Maldacena:1997re} 
  J.~M.~Maldacena,
  ``The Large N limit of superconformal field theories and supergravity,''
  Adv.\ Theor.\ Math.\ Phys.\  {\bf 2}, 231 (1998)
  [hep-th/9711200].
\bibitem{Witten:1998qj} 
  E.~Witten,
  ``Anti-de Sitter space and holography,''
  Adv.\ Theor.\ Math.\ Phys.\  {\bf 2}, 253 (1998)
  [hep-th/9802150].
\bibitem{Witten:1998zw} 
  E.~Witten,
  ``Anti-de Sitter space, thermal phase transition, and confinement in gauge theories,''
  Adv.\ Theor.\ Math.\ Phys.\  {\bf 2}, 505 (1998)
  [hep-th/9803131].
\bibitem{Gubser:1998bc} 
  S.~S.~Gubser, I.~R.~Klebanov and A.~M.~Polyakov,
  ``Gauge theory correlators from noncritical string theory,''
  Phys.\ Lett.\ B {\bf 428}, 105 (1998)
  [hep-th/9802109].
\bibitem{Karch:2006pv} 
  A.~Karch, E.~Katz, D.~T.~Son and M.~A.~Stephanov,
  ``Linear confinement and AdS/QCD,''
  Phys.\ Rev.\ D {\bf 74}, 015005 (2006)
  [hep-ph/0602229].
\bibitem{Li:2013oda}
D.~Li and M.~Huang,
``Dynamical holographic QCD model for glueball and light meson spectra,''
JHEP \textbf{11}, 088 (2013)
[arXiv:1303.6929 [hep-ph]].
\bibitem{Capossoli:2015ywa}
E.~Folco Capossoli and H.~Boschi-Filho,
``Glueball spectra and Regge trajectories from a modified holographic softwall model,''
Phys. Lett. B \textbf{753}, 419-423 (2016)
[arXiv:1510.03372 [hep-ph]].
\bibitem{BallonBayona:2007qr} 
  C.~A.~Ballon Bayona, H.~Boschi-Filho and N.~R.~F.~Braga,
  ``Deep inelastic scattering from gauge string duality in the soft wall model,''
  JHEP {\bf 0803}, 064 (2008)
  [arXiv:0711.0221 [hep-th]].
\bibitem{Braga:2011wa} 
  N.~R.~F.~Braga and A.~Vega,
  ``Deep inelastic scattering of baryons in a modified soft wall model,''
  Eur.\ Phys.\ J.\ C {\bf 72}, 2236 (2012)
  [arXiv:1110.2548 [hep-ph]].

\bibitem{Huang:2007fv} 
  S.~He, M.~Huang, Q.~S.~Yan and Y.~Yang,
  ``Confront Holographic QCD with Regge Trajectories,''
  Eur.\ Phys.\ J.\ C {\bf 66}, 187 (2010)
  [arXiv:0710.0988 [hep-ph]].
\bibitem{Branz:2010ub} 
  T.~Branz, T.~Gutsche, V.~E.~Lyubovitskij, I.~Schmidt and A.~Vega,
  ``Light and heavy mesons in a soft-wall holographic approach,''
  Phys.\ Rev.\ D {\bf 82}, 074022 (2010)
  [arXiv:1008.0268 [hep-ph]].
\bibitem{Gutsche:2011vb} 
  T.~Gutsche, V.~E.~Lyubovitskij, I.~Schmidt and A.~Vega,
  ``Dilaton in a soft-wall holographic approach to mesons and baryons,''
  Phys.\ Rev.\ D {\bf 85}, 076003 (2012)
  [arXiv:1108.0346 [hep-ph]].
\bibitem{Afonin:2012jn} 
  S.~S.~Afonin,
  ``Generalized Soft Wall Model,''
  Phys.\ Lett.\ B {\bf 719}, 399 (2013)
  [arXiv:1210.5210 [hep-ph]].
\bibitem{Fang:2016uer} 
  Z.~Fang, D.~Li and Y.~L.~Wu,
  ``IR-improved Soft-wall AdS/QCD Model for Baryons,''
  Phys.\ Lett.\ B {\bf 754}, 343 (2016)
  [arXiv:1602.00379 [hep-ph]].
\bibitem{Cortes:2017lgz} 
  S.~Cortés, M.~Á.~Martin ~Contreras and J.~R.~Roldán,
  ``Light Meson Masses using AdS/QCD modified Soft Wall Model,''
  Phys.\ Rev.\ D {\bf 96}, no. 10, 106002 (2017)
  [arXiv:1706.09502 [hep-ph]].

\bibitem{Afonin:2018era} 
  S.~S.~Afonin and A.~D.~Katanaeva,
  ``Glueballs and deconfinement temperature in AdS/QCD,''
  Phys.\ Rev.\ D {\bf 98}, no. 11, 114027 (2018)
  [arXiv:1809.07730 [hep-ph]].
\bibitem{Gutsche:2019blp}
T.~Gutsche, V.~E.~Lyubovitskij, I.~Schmidt and A.~Y.~Trifonov,
``Mesons in a soft-wall AdS-Schwarzschild approach at low temperature,''
Phys. Rev. D \textbf{99}, no.5, 054030 (2019)
[arXiv:1902.01312 [hep-ph]].
\bibitem{Contreras:2018hbi}
M.~\'A.~Mart\'\i{}n Contreras, A.~Vega and S.~Cort\'es,
``Light pseudoscalar and axial spectroscopy using AdS/QCD modified soft wall model,''
Chin. J. Phys. \textbf{66}, 715-723 (2020)
[arXiv:1811.10731 [hep-ph]].
\bibitem{Andreev:2006vy} 
  O.~Andreev,
  ``1/q**2 corrections and gauge/string duality,''
  Phys.\ Rev.\ D {\bf 73}, 107901 (2006)
  [hep-th/0603170].
\bibitem{Andreev:2006ct} 
  O.~Andreev and V.~I.~Zakharov,
  ``Heavy-quark potentials and AdS/QCD,''
  Phys.\ Rev.\ D {\bf 74}, 025023 (2006)
  [hep-ph/0604204].
\bibitem{FolcoCapossoli:2019imm}
E.~Folco Capossoli, M.~A.~M.~Contreras, D.~Li, A.~Vega and H.~Boschi-Filho,
``Hadronic Spectra from Deformed AdS Backgrounds,''
Chin. Phys. C \textbf{44}, no.6, 064104 (2020)
[arXiv:1903.06269 [hep-ph]].
\bibitem{Forkel:2007cm} 
  H.~Forkel, M.~Beyer and T.~Frederico,
  ``Linear square-mass trajectories of radially and orbitally excited hadrons in holographic QCD,''
  JHEP {\bf 0707}, 077 (2007)
  [arXiv:0705.1857 [hep-ph]].
\bibitem{White:2007tu} 
  C.~D.~White,
  ``The Cornell potential from general geometries in AdS / QCD,''
  Phys.\ Lett.\ B {\bf 652}, 79 (2007)
  [hep-ph/0701157].
\bibitem{Wang:2009wx}
C.~Wang, S.~He, M.~Huang, Q.~S.~Yan and Y.~Yang,
``Scalar Mesons and glueballs in Dp-Dq hard-wall models,''
Chin. Phys. C \textbf{34}, 319-324 (2010)
[arXiv:0902.0864 [hep-ph]].
\bibitem{Rinaldi:2017wdn} 
  M.~Rinaldi and V.~Vento,
  ``Scalar and Tensor Glueballs as Gravitons,''
  Eur.\ Phys.\ J.\ A {\bf 54}, 151 (2018)
  [arXiv:1710.09225 [hep-ph]].
\bibitem{Bruni:2018dqm} 
  R.~C.~L.~Bruni, E.~Folco Capossoli and H.~Boschi-Filho,
  ``Quark-antiquark potential from a deformed AdS/QCD,''
  Adv.\ High Energy Phys.\  {\bf 2019}, 1901659 (2019)
  [arXiv:1806.05720 [hep-th]].
\bibitem{Diles:2018wbe}
S.~Diles,
``Probing AdS/QCD backgrounds with semi-classical strings,''
[arXiv:1811.03141 [hep-th]].
\bibitem{Tahery:2020tub}
S.~Tahery and X.~Chen,
``Drag force on a moving heavy quark with deformed string configuration,''
[arXiv:2004.12056 [hep-th]].
\bibitem{Caldeira:2020sot}
N.~G.~Caldeira, E.~Folco Capossoli, C.~A.~D.~Zarro and H.~Boschi-Filho,
``Fluctuation and dissipation from a deformed string/gauge duality model,''
Phys. Rev. D \textbf{102}, no.8, 086005 (2020)
[arXiv:2007.00160 [hep-th]].
\bibitem{Chen:2020ath}
X.~Chen, L.~Zhang, D.~Li, D.~Hou and M.~Huang,
``Gluodynamics and deconfinement phase transition under rotation from holography,''
[arXiv:2010.14478 [hep-ph]].
\bibitem{Caldeira:2020rir}
N.~G.~Caldeira, E.~Folco Capossoli, C.~A.~D.~Zarro and H.~Boschi-Filho,
``Fluctuation and dissipation within a deformed holographic model with backreaction,''
Phys. Lett. B \textbf{815}, 136140 (2021)
[arXiv:2010.15293 [hep-th]].
\bibitem{Rinaldi:2021dxh}
M.~Rinaldi and V.~Vento,
``Meson and glueball spectroscopy within the graviton soft wall model,''
[arXiv:2101.02616 [hep-ph]].
\cite{Grigoryan:2007wn}
\bibitem{Grigoryan:2007wn}
H.~R.~Grigoryan and A.~V.~Radyushkin,
``Pion form-factor in chiral limit of hard-wall AdS/QCD model,''
Phys. Rev. D \textbf{76}, 115007 (2007)
[arXiv:0709.0500 [hep-ph]].
\bibitem{Ferreira:2019inu}
L.~F.~Ferreira and R.~Da Rocha,
``Pion family in AdS/QCD: the next generation from configurational entropy,''
Phys. Rev. D \textbf{99}, no.8, 086001 (2019)
[arXiv:1902.04534 [hep-th]].
\bibitem{Lv:2018wfq}
M.~Lv, D.~Li and S.~He,
``Pion condensation in a soft-wall AdS/QCD model,''
JHEP \textbf{11}, 026 (2019)
[arXiv:1811.03828 [hep-ph]].
\bibitem{Bacchetta:2017vzh}
A.~Bacchetta, S.~Cotogno and B.~Pasquini,
``The transverse structure of the pion in momentum space inspired by the AdS/QCD correspondence,''
Phys. Lett. B \textbf{771}, 546-552 (2017)
[arXiv:1703.07669 [hep-ph]].
\bibitem{Grigoryan:2008up}
H.~R.~Grigoryan and A.~V.~Radyushkin,
``Anomalous Form Factor of the Neutral Pion in Extended AdS/QCD Model with Chern-Simons Term,''
Phys. Rev. D \textbf{77}, 115024 (2008)
[arXiv:0803.1143 [hep-ph]].
\bibitem{Brodsky:2007hb}
S.~J.~Brodsky and G.~F.~de Teramond,
``Light-Front Dynamics and AdS/QCD Correspondence: The Pion Form Factor in the Space- and Time-Like Regions,''
Phys. Rev. D \textbf{77}, 056007 (2008)
[arXiv:0707.3859 [hep-ph]].
\bibitem{Kwee:2007dd}
H.~J.~Kwee and R.~F.~Lebed,
``Pion form-factors in holographic QCD,''
JHEP \textbf{01}, 027 (2008)
[arXiv:0708.4054 [hep-ph]].
\bibitem{Vega:2008te}
A.~Vega and I.~Schmidt,
``Hadrons in AdS/QCD correspondence,''
Phys. Rev. D \textbf{79}, 055003 (2009)
[arXiv:0811.4638 [hep-ph]].
\bibitem{Bayona:2010bg}
C.~A.~B.~Bayona, H.~Boschi-Filho, M.~Ihl and M.~A.~C.~Torres,
``Pion and Vector Meson Form Factors in the Kuperstein-Sonnenschein holographic model,''
JHEP \textbf{08}, 122 (2010)
[arXiv:1006.2363 [hep-th]].
\bibitem{Zuo:2009hz}
F.~Zuo, Y.~Jia and T.~Huang,
``gamma* rho0 ---\ensuremath{>} pi0 Transition Form Factor in Extended AdS/QCD Models,''
Eur. Phys. J. C \textbf{67}, 253-261 (2010)
[arXiv:0910.3990 [hep-ph]].
\bibitem{Stoffers:2011xe}
A.~Stoffers and I.~Zahed,
``$\gamma^* gamma^* \to \pi^0$ Form Factor from AdS/QCD,''
Phys. Rev. C \textbf{84}, 025202 (2011)
[arXiv:1104.2081 [hep-ph]].
\bibitem{Gutsche:2014zua}
T.~Gutsche, V.~E.~Lyubovitskij, I.~Schmidt and A.~Vega,
``Pion light-front wave function, parton distribution and the electromagnetic form factor,''
J. Phys. G \textbf{42}, no.9, 095005 (2015)
[arXiv:1410.6424 [hep-ph]].
\bibitem{BoschiFilho:2002ta}
H.~Boschi-Filho and N.~R.~F.~Braga,
``QCD / string holographic mapping and glueball mass spectrum,''
Eur. Phys. J. C \textbf{32}, 529-533 (2004)
[arXiv:hep-th/0209080 [hep-th]].
\bibitem{BoschiFilho:2002vd}
H.~Boschi-Filho and N.~R.~F.~Braga,
``Gauge / string duality and scalar glueball mass ratios,''
JHEP \textbf{05}, 009 (2003)
[arXiv:hep-th/0212207 [hep-th]].
\bibitem{Polchinski:2002jw} 
  J.~Polchinski and M.~J.~Strassler,
  ``Deep inelastic scattering and gauge / string duality,''
  JHEP {\bf 0305}, 012 (2003)
  [hep-th/0209211].
\bibitem{FolcoCapossoli:2020pks}
E.~Folco Capossoli, M.~A.~Mart\'\i{}n Contreras, D.~Li, A.~Vega and H.~Boschi-Filho,
``Proton structure functions from an AdS/QCD model with a deformed background,''
Phys. Rev. D \textbf{102}, no.8, 086004 (2020)
[arXiv:2007.09283 [hep-ph]].
\bibitem{pdg}
P.A. Zyla et al. (Particle Data Group), ``Prog. Theor. Exp. Phys. 2020, 083C01 (2020)".
\bibitem{Hatta:2007he} 
  Y.~Hatta, E.~Iancu and A.~H.~Mueller,
  ``Deep inelastic scattering at strong coupling from gauge/string duality: The Saturation line,''
  JHEP {\bf 0801}, 026 (2008)
  [arXiv:0710.2148 [hep-th]].
\bibitem{Capossoli:2015sfa} 
  E.~Folco Capossoli and H.~Boschi-Filho,
  ``Deep Inelastic Scattering in the Exponentially Small Bjorken Parameter Regime from the Holographic Softwall Model,''
  Phys.\ Rev.\ D {\bf 92}, no. 12, 126012 (2015)
  [arXiv:1509.01761 [hep-th]].

\bibitem{abramowitz}
M. Abramowitz and I. A. Stegun, \emph{Handbook of Mathematical Functions} (Dover, New York, 1972).
\bibitem{Polchinski:2001ju}
J.~Polchinski and L.~Susskind,
``String theory and the size of hadrons,''
[arXiv:hep-th/0112204 [hep-th]].

\bibitem{Brauel:1979zk}
P.~Brauel, T.~Canzler, D.~Cords, R.~Felst, G.~Grindhammer, M.~Helm, W.~D.~Kollmann, H.~Krehbiel and M.~Schadlich,
``Electroproduction of $\pi^+ n$, $\pi^- p$ and $K^+ \Lambda$, $K^+ \Sigma^0$ Final States Above the Resonance Region,''
Z. Phys. C \textbf{3}, 101 (1979)

\bibitem{Horn:2006tm}
T.~Horn \textit{et al.} [Jefferson Lab F(pi)-2],
``Determination of the Charged Pion Form Factor at Q**2 = 1.60 and 2.45-(GeV/c)**2,''
Phys. Rev. Lett. \textbf{97}, 192001 (2006)
[arXiv:nucl-ex/0607005 [nucl-ex]].



\bibitem{Brodsky:1973kr}
S.~J.~Brodsky and G.~R.~Farrar,
``Scaling Laws at Large Transverse Momentum,''
Phys. Rev. Lett. \textbf{31}, 1153-1156 (1973)
\bibitem{Matveev:1973ra}
V.~A.~Matveev, R.~M.~Muradian and A.~N.~Tavkhelidze,
``Automodellism in the large - angle elastic scattering and structure of hadrons,''
Lett. Nuovo Cim. \textbf{7}, 719-723 (1973)
\bibitem{Efremov:2009dx}
A.~Efremov and A.~Radyushkin,
``Perturbative QCD of Hard and Soft Processes,''
Mod. Phys. Lett. A \textbf{24}, 2803-2824 (2009)
[arXiv:0911.1195 [hep-ph]].
\bibitem{Colangelo:2008us}
P.~Colangelo, F.~De Fazio, F.~Giannuzzi, F.~Jugeau and S.~Nicotri,
``Light scalar mesons in the soft-wall model of AdS/QCD,''
Phys. Rev. D \textbf{78} (2008), 055009
[arXiv:0807.1054 [hep-ph]].

\bibitem{Grigoryan:2007my}
H.~R.~Grigoryan and A.~V.~Radyushkin,
``Structure of vector mesons in holographic model with linear confinement,''
Phys. Rev. D \textbf{76} (2007), 095007
[arXiv:0706.1543 [hep-ph]].





















































































\end{thebibliography}
\end{document}